\documentclass[14pt, preprint2]{aastex6}
\usepackage{bm, graphicx, subfigure, amsmath, morefloats, mathtools}
\hypersetup{allcolors = [rgb]{0., 0., 0.5}} 

\usepackage{longtable}
\newcommand{\project}[1]{\textsl{#1}}
 
\newcommand{\tc}{\project{The~Cannon}} 
\newcommand{\apogee}{\project{\textsc{apogee}}}

\newcommand{\aspcap}{\project{\textsc{aspcap}}}

\newcommand{\teff}{\mbox{$\rm T_{eff}$}}

\newcommand{\feh}{\mbox{$\rm [Fe/H]$}}

\newcommand{\logg}{\mbox{$\rm \log g$}}

\newcommand{\given}{\,|\,}

\newcommand{\xone}{\ensuremath{x_n}}
\newcommand{\xtwo}{\ensuremath{x_{n'}}}
\newcommand{\done}{\ensuremath{\sigma_n}}
\newcommand{\dtwo}{\ensuremath{\sigma_{n'}}}
\newcommand{\xbar}{\ensuremath{\overline{x}}}


\begin{document}

\title{Galactic doppelganger: The chemical similarity among field stars and among stars with a common birth origin}
\author{M.~Ness\altaffilmark{1},
        H-W.~Rix\altaffilmark{1}, 
    David~W.~Hogg\altaffilmark{1,2,3,4},
         A.R.~Casey\altaffilmark{5},
         J.~Holtzman\altaffilmark{6}, 
      M.~Fouesneau\altaffilmark{1}, 
         G.~Zasowski\altaffilmark{7,8},
         D.~Geisler\altaffilmark{9},
         M.~Shetrone\altaffilmark{10},
         D.~Minniti\altaffilmark{11,12,13}
         Peter M.~Frinchaboy\altaffilmark{14},
         Alexandre~Roman-Lopes\altaffilmark{15}}

\altaffiltext{1}{Max-Planck-Institut f\"ur Astronomie, K\"onigstuhl 17, D-69117 Heidelberg, Germany}
\altaffiltext{2}{Center for Cosmology and Particle Physics, Department of Physics, New York University, 726 Broadway, New York, NY 10003}
\altaffiltext{3}{Center for Data Science, New York University, 60 5th Avenue, New York, NY 10011, USA}
\altaffiltext{4}{Flatiron Institute, Simons Foundation, 162 Fifth Ave, New York, NY 10010}
\altaffiltext{5}{Institute of Astronomy, University of Cambridge,  Madingley Road, Cambridge CB3~0HA, UK}
\altaffiltext{6}{Department of Astronomy, New Mexico State University, Las Cruces, NM 88003, USA}
\altaffiltext{7}{Department of Physics \& Astronomy, University of Utah, 115 S. 1400 E., Salt Lake City, UT 84112, USA}
\altaffiltext{8}{Space Telescope Science Institute, 3700 San Martin Drive, Baltimore MD 21218, USA}
\altaffiltext{9}{Departamento de Astronomia, Casilla 160-C, 
       Universidad de Concepcion, Chile}
\altaffiltext{10}{University of Texas at Austin, McDonald Observatory, USA}
       
\altaffiltext{11}{Instituto Milenio de Astrofisica, Santiago, Chile}
\altaffiltext{12}{Departamento de Fisica, Facultad de Ciencias Exactas, Universidad Andres Bello, 
Av. Fernandez Concha 700, Las Condes, Santiago, Chile}
\altaffiltext{13}{Vatican Observatory, V00120 Vatican City State, Italy}
\altaffiltext{14}{Department of Physics \& Astronomy, Texas Christian
University,
TCU Box 298840, Fort Worth, TX 76129, USA}
\altaffiltext{15}{Department of Physics \& Astronomy - Universidad de La Serena - A. Juan Cisternas, 1200 North, La Serena, Chile}

\email{ness@mpia.de}

\begin{abstract} 
We explore to which extent stars within Galactic disk open clusters resemble each other in the high-dimensional space of their photospheric element abundances, and contrast this with pairs of field stars. Our analysis is based on abundances for 20 elements, homogeneously derived from \apogee\ spectra (with carefully quantified uncertainties, with a median value of $\sim 0.03$ dex). We consider 90 red giant stars in seven open clusters and find that {\it most} stars within a cluster have abundances in {\it most} elements that are indistinguishable  (in a $\chi^2$-sense) from those of the other members, as expected for stellar birth siblings. An analogous analysis among pairs of $>1000$ field stars shows that highly significant abundance differences in the 20-dimensional space can be established for the vast majority of these pairs, and that the \apogee -based abundance measurements have high discriminating power. However, pairs of field stars whose abundances are indistinguishable even at 0.03~dex precision exist:  $\sim 0.3$\% of all field star pairs, and $\sim 1.0$\% of field star pairs at the same (solar) metallicity [Fe/H] = $0 \pm 0.02$. Most of these pairs are presumably not birth siblings from the same cluster, but rather {\it doppelganger}. Our analysis implies that “chemical tagging” in the strict sense, identifying birth siblings for typical disk stars through their abundance similarity alone, will not work with such data. However, our approach shows that abundances have extremely valuable information for probabilistic chemo-orbital modeling and combined with velocities, we have identified new cluster members from the field.
\end{abstract}

\section{Introduction}

There now exists an abundance of spectroscopic data for stars across the Milky Way, from which large numbers of element abundances have been, or can be, measured. These data are being delivered by current surveys including Gaia-ESO \citep{gilmore2012}, APOGEE \citep{Majewski2016}, RAVE \citep{Kunder2016, Casey2016b} and GALAH \citep{Freeman2012, deSilva2015} and there are numerous future spectroscopic surveys being planned including 4MOST \citep{deJong2015}, MOONS \citep{C2012} and WEAVE \citep{D2012}. Large volumes of data, combined with new techniques to optimally exploit the data and deliver high precision stellar abundances, e.g., \tc\ \citep{Ness2015, Casey2016}, enable us to examine the distribution in abundance space of the stars in the Milky Way disk and halo. This information can then, in principle, be used to constrain the assembly history of our Galaxy. One way to pursue this is to identify stars of common birth sites via their abundance signatures, a process called chemical tagging \citep{freeman2002}. To do this, we need to first know what the spread, if any, is in the abundance measurements of stars that are known to be born together.

Stars in open clusters are believed to be born from single star forming aggregates and therefore these stars should be chemically homogeneous \citep[e.g.][]{deSilva2007,deSilva2009}. The chemical similarity of stars born together has been used to identify members of moving groups or co-natal groups in the Milky Way including using abundances alone \citep[e.g.][]{Majewski2012a, Hogg2016} as well as chemically exceptional groups of stars \citep[e.g.][]{Schiavon2016, Martell2016}. Many of the individual clusters have been studied in numerous studies, including using \apogee\ data \citep[e.g.][]{Souto2016, Cuhna2015, F2013} and the {\it measured} abundance differences within clusters has been demonstrated to be relatively small, on the order of measurement errors themselves \citep{Reddy2012, Reddy2015, Lambert2016, M2014}. Statistically there is a much greater abundance similarity among stars within a cluster, compared to stars from different clusters \citep[e.g.][]{M2013, deSilva2015}. However, although the question of the magnitude of open cluster abundance dispersion has been investigated previously, there has been very little assessment as to the true \textit{intrinsic} dispersion of clusters in their many elements \citep[see also][]{M2013}.

The abundance dispersion that we measure at face value in groups of stars that are born together in clusters (or potentially also in associations), is a combination of the intrinsic dispersion of the cluster and the measurement uncertainties. Quantifying this intrinsic dispersion for clusters is indispensable not only for the prospects of chemical tagging. This is also critical for understanding the chemical abundance distribution and dimensionality of the Milky Way disk and key for being able to determine when stars are likely to be \textit{not} born together. 

A high level of abundance homogeneity (to within $\sim 0.03$~dex) has been demonstrated for three open clusters in \apogee; M67, NGC6819 and NGC2410, from the stellar spectra directly \citep{Bovy2016}: after removing temperature trends, the cluster spectra form one-dimensional sequences. This novel data-driven approach was motivated by the difficulties in obtaining consistent high precision stellar abundance measurements and such is an approach is optimal to achieve high precision limits on the intrinsic cluster abundance dispersions, and so assess homogeneity of a single birth site itself. Only this data-driven approach, which did not provide actual abundance measurements, has managed to place tight constraints on the intrinsic cluster dispersion \citep{Bovy2016}. However, the limitation in such a method is that it does not return absolute abundance measurements and as such the open cluster measurements can not be directly compared with that of the field, which is our aim here. 

With our modifications to the Cannon and carefully selected high-fidelity training set, we achieve high precision and report absolute abundance measurements for 90 open cluster stars.  From these consistent, high-precision abundance measurements across ranges of temperature and gravity, we make an assessment of the multi-element abundance dispersion (or limits thereon), given our thorough characterization of our measurement uncertainties. Our results constitute the largest homogeneous and consistent analysis of the \textit{intrinsic} abundance dispersion within open clusters. 

Comparing these high precision abundance measurements in \apogee\  among stars in an
open cluster and among disk field stars, can shape our expectation for the
chemical similarity of stars that were born together -- and those that were presumably not (but are still members of the Milky Way's dominant disk population). We should expect that -- with abundance data of this quality -- most intra-cluster pairs of stars look very similar or indistinguishable, while most pairs of field stars have discernible abundance differences. Importantly, these data can allow us to determine the fraction of field stars that are not related in birth origin, yet are as chemically similar as those  born in the same molecular cloud. In their abundances, these pairs of stars look perfectly alike (given the 20 \apogee\ abundances), but are in most cases {\it not} birth siblings from the same cluster; we dub those stars {\it doppelganger}. The rate of {\it doppelganger} obviously plays a decisive role in the efficacy of chemical tagging. 

In our analysis, there are two key ingredients for delivering our high precision measurements from APOGEE data: (i) We modified our spectral modeling, \tc, to correct for spurious abundance variations across fiber number, caused by the varying Line Spread Function (LSF) across the APOGEE detectors; and in \tc 's training step we used abundances that have been corrected for LSF variations. (ii) We selected a high fidelity training set using labels from \apogee's stellar parameter and abundances pipeline ASPCAP \citep{GP2016}: 5000 high signal-to-noise stars that made it sensible to train \tc\ on a total of 23 labels:  three stellar parameters, \teff, \logg\ and \feh, 19 [X/Fe] individual abundances from DR13 \citep[][Johnson et al., 2016, submitted]{Holtzman2015}, plus the mean Line Spread Function (LSF) of the star.   In Section 2 we present our training set and in Section 3 we discuss our modifications to \tc\ and training set in order to cope with the problem of mean abundance changing as a function of LSF. In Section 4 we show the precision we achieve for our abundances, from the cross-validation of the training set and report the abundances we obtain for the individual clusters. We use our uncertainties to determine the intrinsic spread in each element using our cluster model. In Section 5 we examine the intra-cluster and field pair abundance similarity distributions and from this: we demonstrate that the abundance similarity among pairs of stars within clusters, defined via a $\chi^2$-distance in 20-element abundance space, is far greater than among random pairs of field stars.  We also determine the doppelganger rate among Galactic disk stars (for data sets of this quality): we find this rate to be very small ($\sim 1\%$), yet large enough to matter greatly for chemical tagging. In Section 6 we present as an aside new cluster members that we have identified from the field using our approach of examining stellar element abundance similarity between pairs. In Section 7 we discuss the implications of our results. 

Our work makes the first quantification of the doppelganger rate, which directly tests the viability of strict chemical tagging. Our analysis also demonstrates the diagnostic power in the high dimensional abundance space: as part of our analysis, we identify new cluster members, by combining the chemical information using our pair similarity measure and radial velocity data.  The high precision measurements that we derive are therefore important for the broader assessment of disk (and cluster) formation. In general, such high precision measurements as we derive, with their carefully characterized uncertainties, are absolutely necessary to make use of the large data sets to characterize the chemical diversity, distribution and, conditioned on expectations of the properties of open clusters, the chemical dimensionality and variability of the Milky Way disk. 

\section{APOGEE Data}

For our analysis we use two aspects of the APOGEE data: we used the \apogee\ spectra \citep{sdssiv, Majewski2012} and stellar parameter and abundances from the SDSS-IV public data release DR13 \citep[][]{Wilson2012, Zasowski2013,  Nidever2015, Holtzman2015, GP2016}. We performed our own signal-to-noise independent continuum normalization on the aspcapStar files, similarly to \citet{Ness2015}, by fitting a second order polynomial to pixels identified with weak parameter dependencies. For all DR13 spectra we selected around 5~percent of the pixels using a criteria of 0.985 $>$ flux $>$ 1.025 and the spectral model coefficients ($|\theta_{Teff}|$, $|\theta_{logg}|$, $|\theta_{[Fe/H]}|$) $<$  (0.005,0.005,0.005). This normalization was applied consistently to the \apogee\ spectra in both the training and the test step of \tc\ \citep{Ness2015}. 

\subsection{Training Data} 

We constructed a training set of data (including their reference labels and their spectra) that consists of 5000 stars with SNR $>$ 200 that span the chemical space of the stars of the Milky Way disk and broadly encompass the label space of the cluster members in the test step.  For our reference labels we used the so called \aspcap\ DR13-corrected labels with small additional corrections for the LSF dependence (see Section 3). We eliminated stars
with highly anomalous abundance measurements and the ASPCAP BAD flag checked.  
Our reference labels span the following range in stellar parameters:\\

\noindent{ 3650 $<$ \teff\ $<$ 5760 K } \\
 0.45 $<$  \logg\ $<$ 3.95 dex \\
--1.7 $<$ \feh\ $<$ 0.36 dex \\

\subsection{Test Data} 

Our test data are the spectra of 97 stars in  seven open clusters, where our membership is taken from the cross over between those identified by \citet{Meszaros2013} and those in the \apogee\ calibration file table \footnote{cal$\_$dr13.fits available at sdss.org}. For NGC2420 we took an additional 3 stars to those identified in \citet{Meszaros2013}  that were studied by \citet{Souto2016}. The properties of the seven open clusters are summarized in Table 1. 

\section{Methodology: Modifications to \tc }

Precise abundance measurements from high SNR spectra are almost inevitably limited by systematics. Systematic trends of abundance with temperature and surface gravity are dealt with in \apogee\ via a post-calibration \citep{Holtzman2015}. The results from this calibration are called DR13-corrected labels. To determine this calibration, the dependence of the abundances on \teff\ and \logg\ is measured for a set of calibration stars, of open and globular cluster stars, plus asteroseismic targets (with very precise \logg\ values derived from the asteroseismic parameters). A polynomial is fit to the \teff\ and \logg\ dependent trends and the results for all \apogee\ stars are adjusted by these fits.  In the course of the present work,
we found a more subtle systematics signature: abundances correlate with the spectrograph's fiber number that the star is assigned to for observation. We attribute this foremost to variations of the LSF across the detector. Potentially, nonuniform characteristics of the electronics, such as the persistence, which affects only part of the (blue) chip from which the spectra are read out (fiber numbers $\leq$ 50) also contribute. 

\begin{figure*}[]
\includegraphics[scale=0.6]{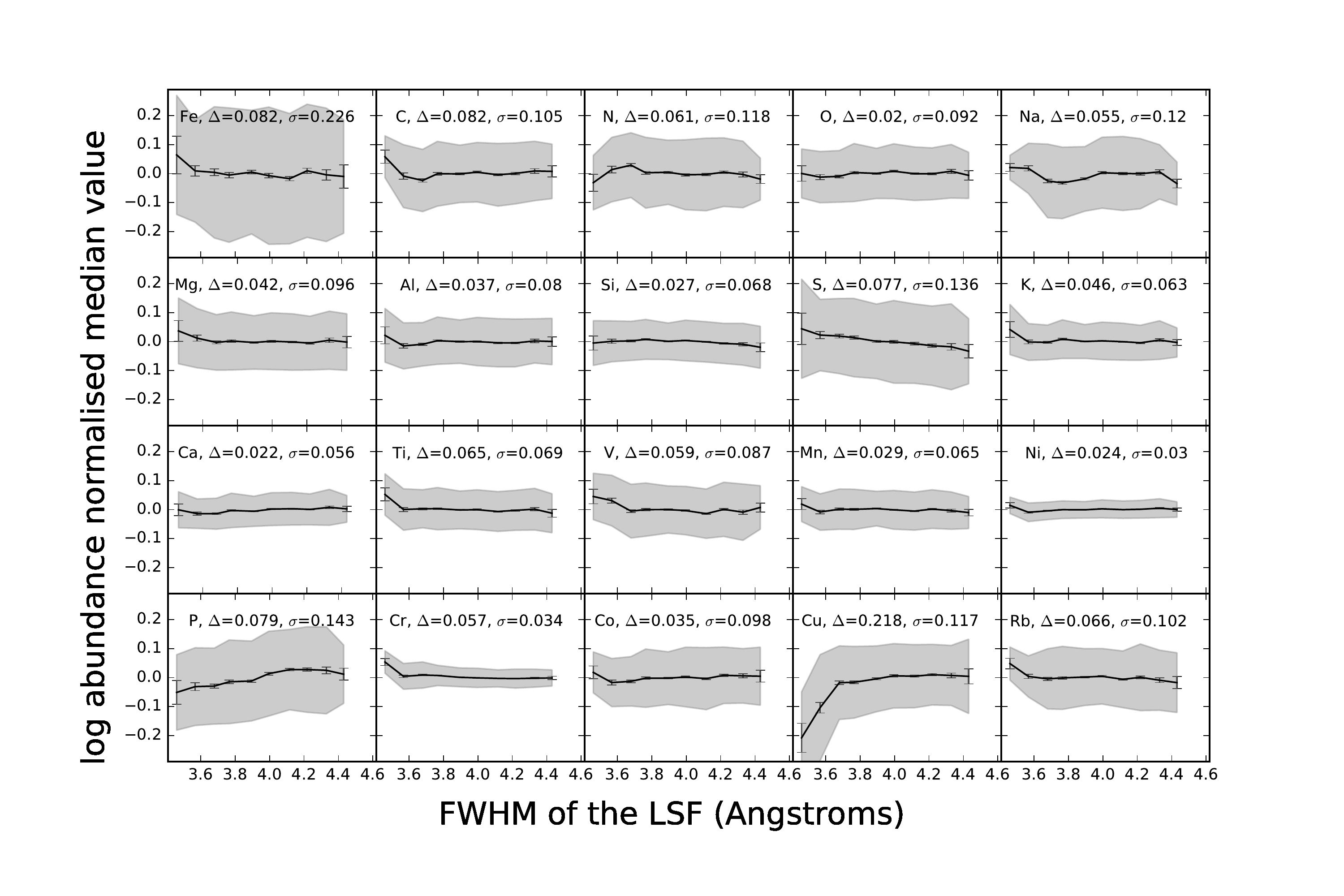} 
  \caption{The normalized median measurements of the \aspcap\ DR13-corrected labels as a function of measured LSF for the 5000 stars used for the training set: the data is normalized by subtracting the overall median measurement from the data, so the data are pivoted around a value = 0, to highlight the offsets on the same scale. The FWHM of the median measurement is shown in the gray shaded region. The largest difference in the median and the dispersion around the median is written at the top of each panel. We correct for these offsets to produce LSF-corrected labels that we use for training for \tc. For many elements this correction is negligible; for some, like P or Cu, it is significant.}
\label{fig:training}
\end{figure*}

Using the DR13 data, we established that the measured abundances depend on LSF width, shown in Figure \ref{fig:training}. This Figure shows the mean DR13 abundances of our training set of stars,  as a function of the Full Width Half Maximum (FWHM) of the LSF for the stars, measured from the apStarLSF files provided by \apogee. For mapping the overall trends of abundances across the disk these trends are not significant and will not affect global measured abundance trends and inferences. For example, with respect to the abundance trends based on the measurements across $(R,z)$, which are on average inclusive of all fiber numbers, independent of position on the sky (except for targeted co-natal groups like streams and open and globular clusters). However, these 
effects may be extremely important when trying to make the most precise abundances comparison among pairs or groups of stars. Therefore, it is necessary to remove these trends before assessing abundance homogeneity between stars. Such effects may increase the abundance differences, but they may also make stars observed with the same fiber appear spuriously similar in their abundances (Section 5). If one takes the DR13 \aspcap\ abundances unaltered, the abundance dispersion of clusters whose stars were observed mostly with low fiber numbers (which equates to the lowest FHWM of the LSF), e.g. NGC2158, NGC6791, M67,  appear higher than those with high fiber number.  This is particularly manifest for ``problematic" elements such as Cu: the mean abundance of which  changes very quickly with LSF or fiber number at the lowest FWHM of the LSF values, which is equivalent to the lowest fiber numbers (Figure \ref{fig:training}). 

To correct for the LSF variation we need to add two processing steps to \tc;  the first affects the reference labels, the second \tc's spectral model.
First, we correct the systematic biases in the reference labels shown in Figure \ref{fig:training} by fitting a 5$^{th}$ order polynomial to the mean trend of each element and adjusting each element by this fit value (given the FWHM of the LSF of the spectrum under consideration).  For most elements this correction is very small or negligible. But for elements like Cu it can be as high as 0.2 dex for the lowest LSF value (see Figure \ref{fig:training}).

Second, we include the FWHM of the LSF as a label in \tc, similarly to the stellar parameters and abundances and we include the FWHM of the LSF as a data point for our model for each star, which is treated in \tc's model in exactly the same way as the flux.  We can consider the LSF's width as a data point, as we know mean LSF width from the apStarLSF files produced for each spectrum. To determine the LSF, we take the average value of the standard deviation of the Gaussian fit across every pixel of the apStarLSF array. This approach to handling LSF variations is an approximation to the ideal approach, which would be to treat the LSF width as deterministically known at both train and test time, and treat it as a convolution and deconvolution operation. That, however, would significantly complexify both the training and test steps, without manifest practical advantage.

\subsubsection{Cross-Validation of this Revised Data Analysis}

\begin{figure*}
\includegraphics[scale=0.46]{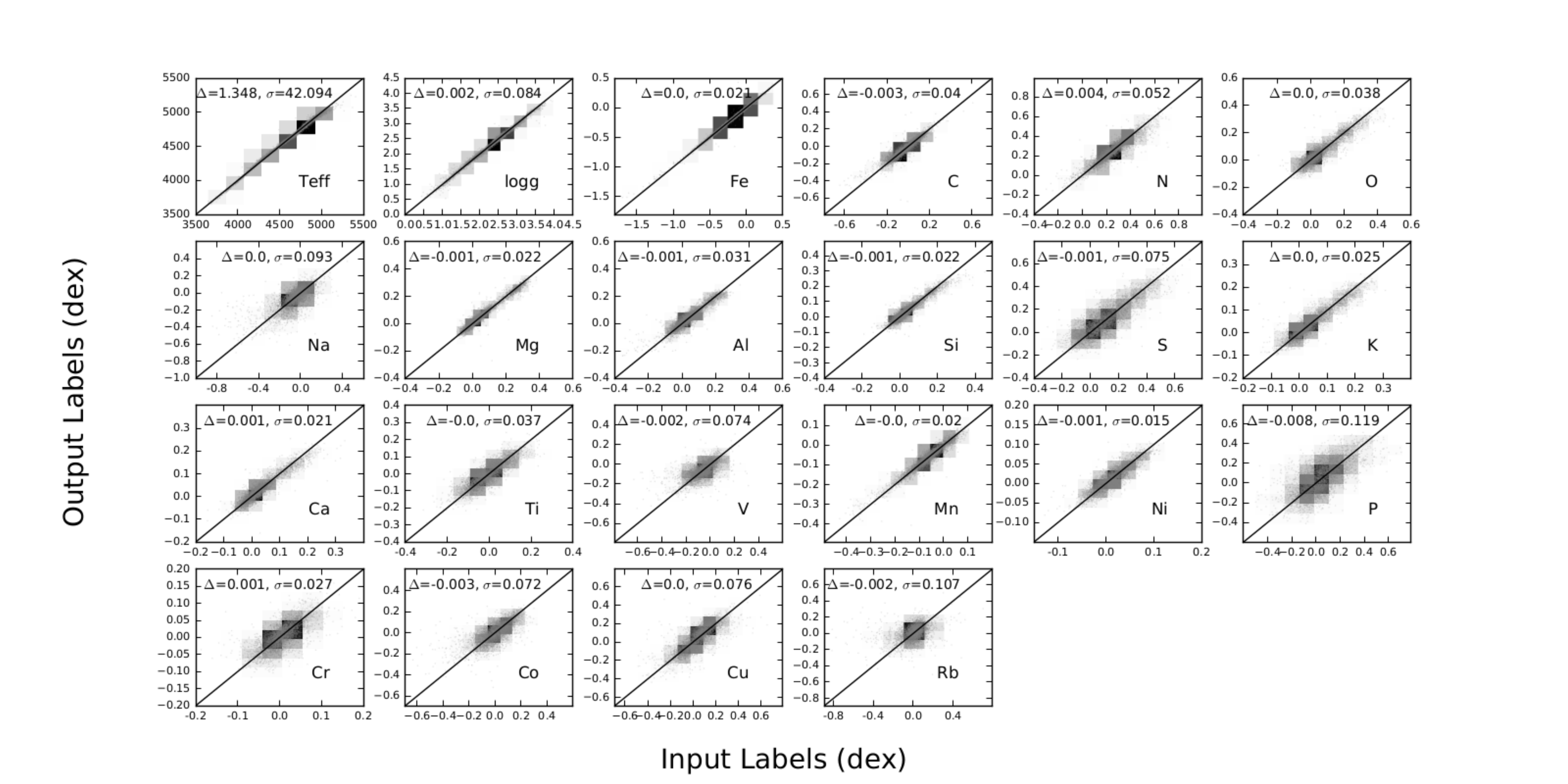} 
  \caption{Abundance cross-validation test, repeatedly removing 10\% of the sample from the training step. All element abundances are with respect to Fe, [X/Fe], except for Fe, which is [Fe/H]. The x-axis shows our input labels (for stars not involved in the training step) and the y-axis shows \tc's output labels. The bias and root mean square deviation for each element ($\Delta$, $\sigma$) are given in each panel. }
\label{fig:cross}
\end{figure*}

Given the slight modifications to \tc\ and the particular reference set, we initiate the analysis with an assessment of the label
precision. We do this with a cross-validation test on the training set, similarly to \citet{Ness2015, Ho2016, Casey2016}. We perform a set of ten leave-10\%-out cross-validations, which are illustrated in Figure \ref{fig:cross}. The stars whose label estimates from \tc\ are shown in this Figure had all been excluded from the training set. The model was constructed using the remaining 90\% of the training set in each case. The x-axis shows the input training labels and the y-axis shows \tc's best fit labels. We are able to recover the labels to high precision: the stellar parameter precisions are $\lesssim$ 45 K in \teff, $\lesssim$ 0.1 dex in \logg\ and $\lesssim$ 0.02 dex in \feh. For individual abundances, the precision is 0.02 to 0.12 dex, depending on the element. 

\begin{figure*}
\centering
           \includegraphics[scale=0.5]{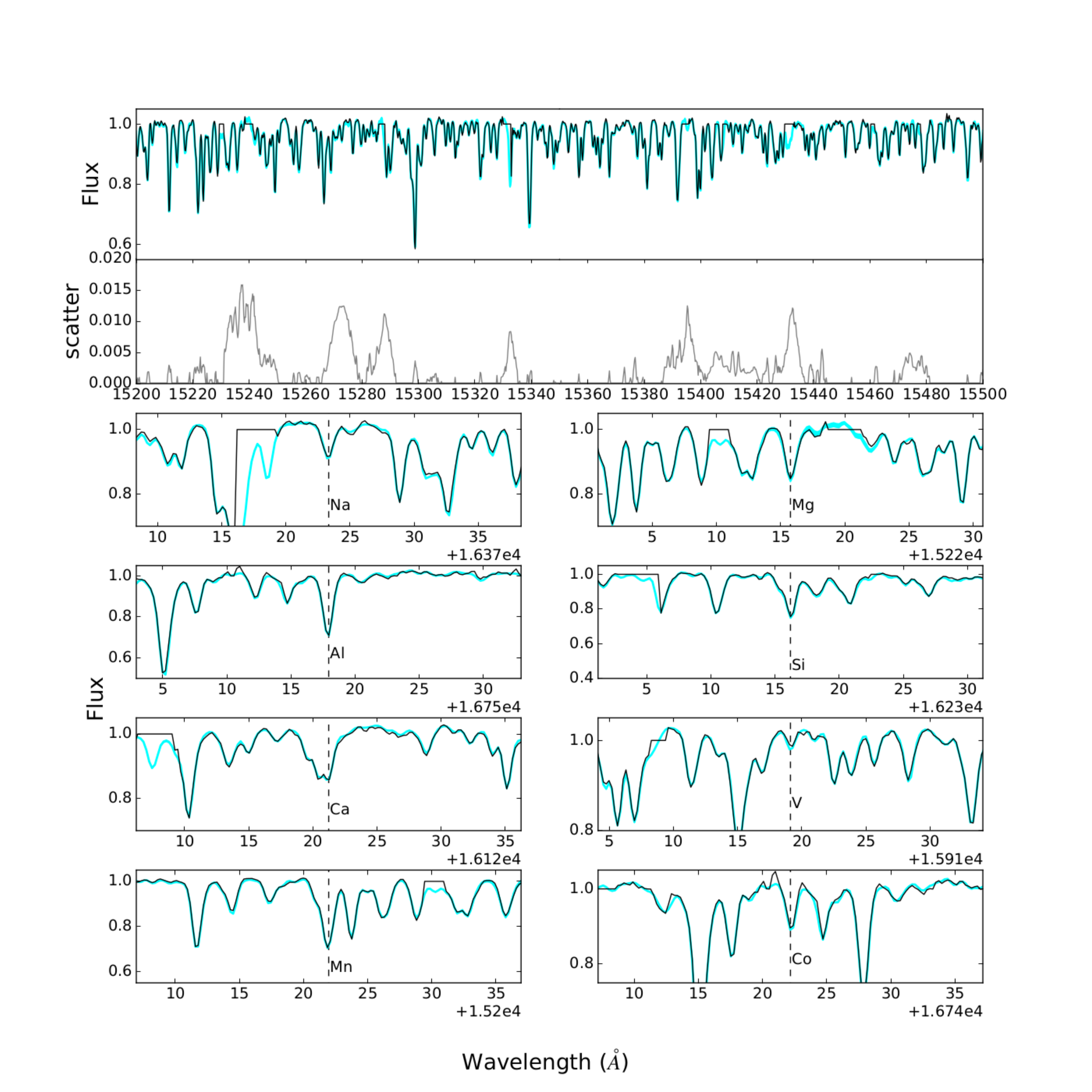}
  \caption{ Example of one star in the cluster NGC6791 with SNR = 260, with the model in cyan and the data in black, showing that the model is a good fit to the data. The top panel shows a 300 Angstrom wavelength range and the second panel shows the model scatter across this region. The lower panels show 30 Angstrom ranges of spectra zoomed in around lines of individual elements, highlighting the goodness of fit to the individual element absorption features. }
\label{fig:typical}
\end{figure*}

Our model fit of \tc\ to the test data is excellent in a $\chi^2$-sense and we show a typical star, from the cluster NGC6791 in Figure \ref{fig:typical}. A  300\AA\  span of the spectrum of this star (in black) and model from \tc\ (in cyan) is shown in the top panel of the Figure. The pixel-by-pixel scatter term of \tc\  is shown in the panel directly underneath, for the same 300\AA\ spectral region. The small scatter quantifies that our model is a good fit to the training data \citep{Ness2015}. Narrow wavelength regions ($\approx$ 30 Angstroms) are also shown in Figure \ref{fig:typical}, to demonstrate the goodness of fit of the model around a number of individual element absorption features. Again, for our model and subsequent derivation of labels, we do not restrict the pixels that \tc\ uses to deliver the abundance information; the cross-validation demonstrates that \tc\ ``learns" where the information about each element is derived from in the spectra, such that the input labels are successfully reproduced at test time.

\section{Results}

We now present the results from these data in three steps. 
First, for an analysis of abundance dispersions, or differences, not only precise
measurements matter, but also a good understanding of the uncertainties: 
over- or underestimates of the individual abundance uncertainties would lead to
under- or overestimates of the cluster-intrinsic abundance dispersions, 
or the abundance differences among pairs of stars. Second, we proceed to quantify the 
abundance dispersions in each of the elements in each of our clusters. 
Third, we show how to characterize the abundance similarities, or differences, 
among pairs of stars; we then apply this to pairs of stars within a cluster 
and within random stars from the field, which of course speaks immediately to the question of chemical tagging.

\begin{figure*}
\centering
\includegraphics[scale=0.35]{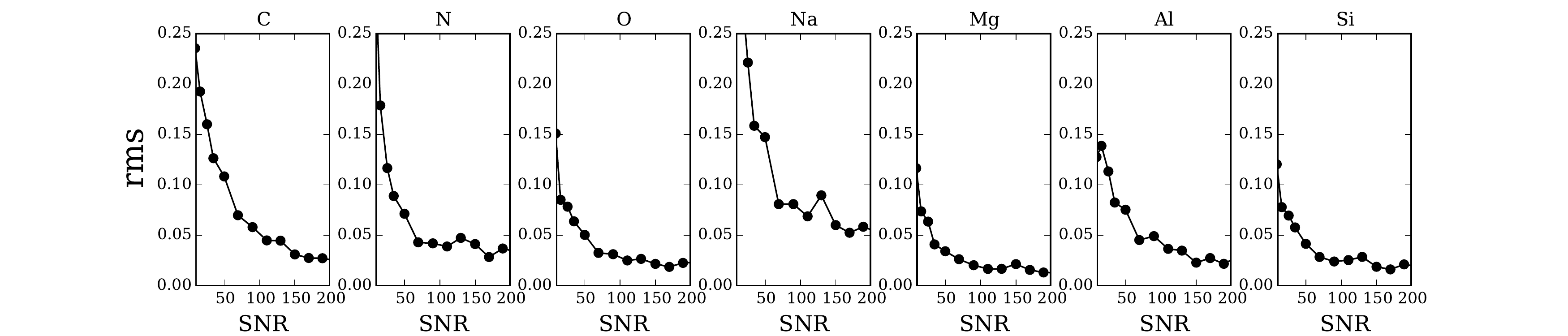} 
  \caption{SNR dependent performance for a sample of the elements, showing the measured rms difference of individual visit measurements on the y-axis. All elements are with respect to Fe, except \feh.}
\label{fig:snr_error}
\end{figure*}

\subsection{Revised APOGEE Abundances and Abundance Uncertainties for Members of the Open Clusters} 

Based on the goodness-of-fit $\chi^2$ metric determined above for each star, we excluded seven stars from the cluster NGC6791, with a $\chi^2$ $>$ 22,000. For our analysis we do not exclude cluster stars with relatively low SNR measurements (SNR $<$ 100), but use the SNR-dependent error to estimate the scaled precision at low SNR. We determine the SNR dependence of the precision using a calibration set of data available as part of the public release of DR13, which includes both co-added spectra and individual visit spectra for a set of $\approx$ 1000 open cluster, globular cluster and calibration stars. A sample of elements is shown in Figure \ref{fig:snr_error}, where each data point is determined by measuring the variance of the difference between the measured abundance from high SNR spectra (comprised of combined individual observations) and the individual spectra with low SNR, for each of the 1000 stars in the calibration set. The Figure shows the dependence across SNR of 10 to 200, where the precision flattens above SNR of about 150. 

Our cluster stars have a reported SNR from 60 to 1000. We the adopt abundance uncertainties for each star that are the quadrature sum of the signal-to-noise scaled uncertainties estimated from the cross-validation and the formal errors returned by the \tc 's test step optimizer.

\begin{figure*}
\centering
\includegraphics[scale=0.5]{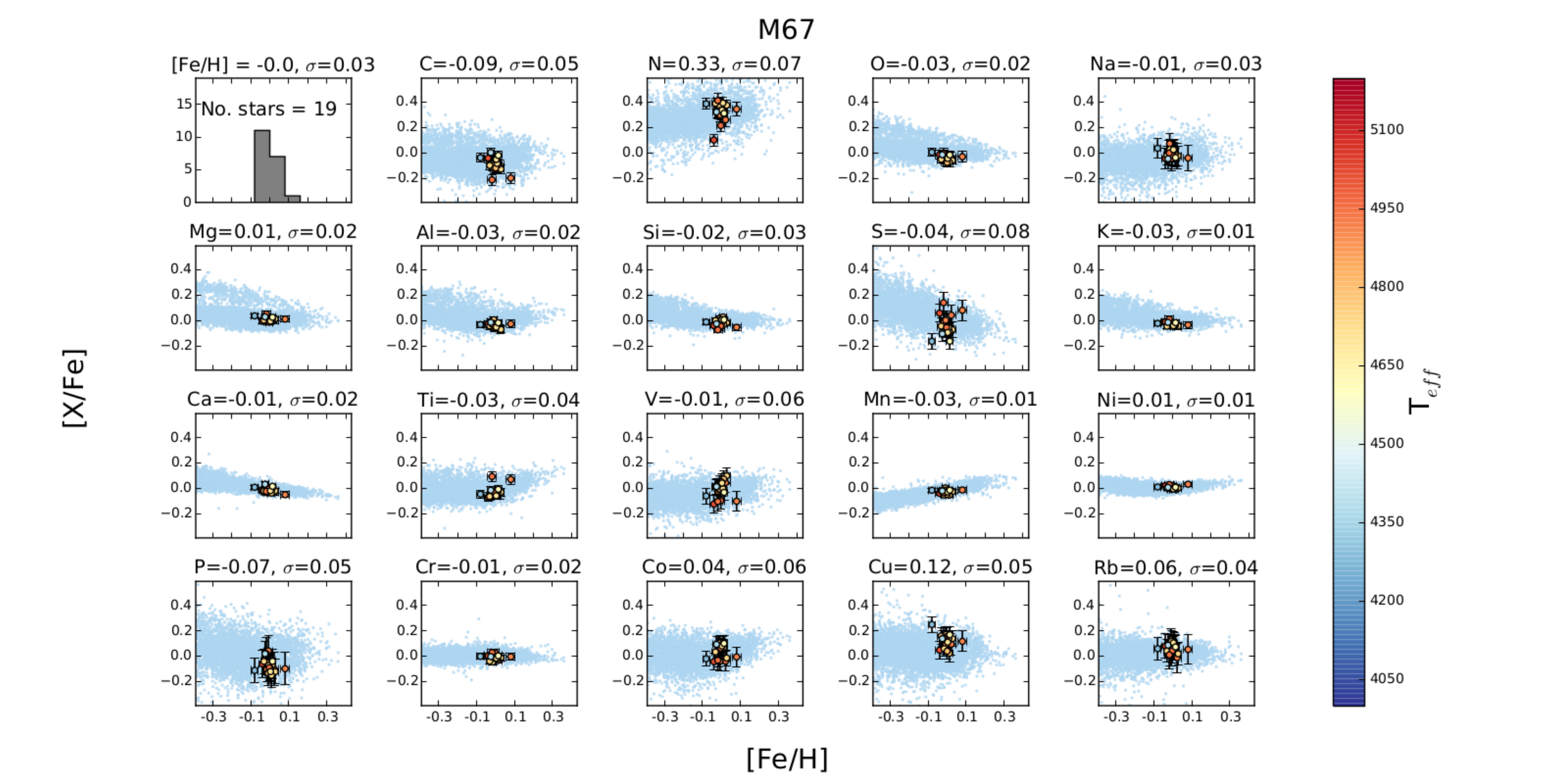}
  \caption{ M67 stars with a median SNR = 438  coloured by effective temperature. The grey points show the training data. The two value at the top of each subplot are the mean and standard deviation of the measurements, where all elements are with respect to Fe except [Fe/H]. }
\label{fig:c1}
\end{figure*}

\begin{figure*}
\centering 
  \includegraphics[scale=0.5]{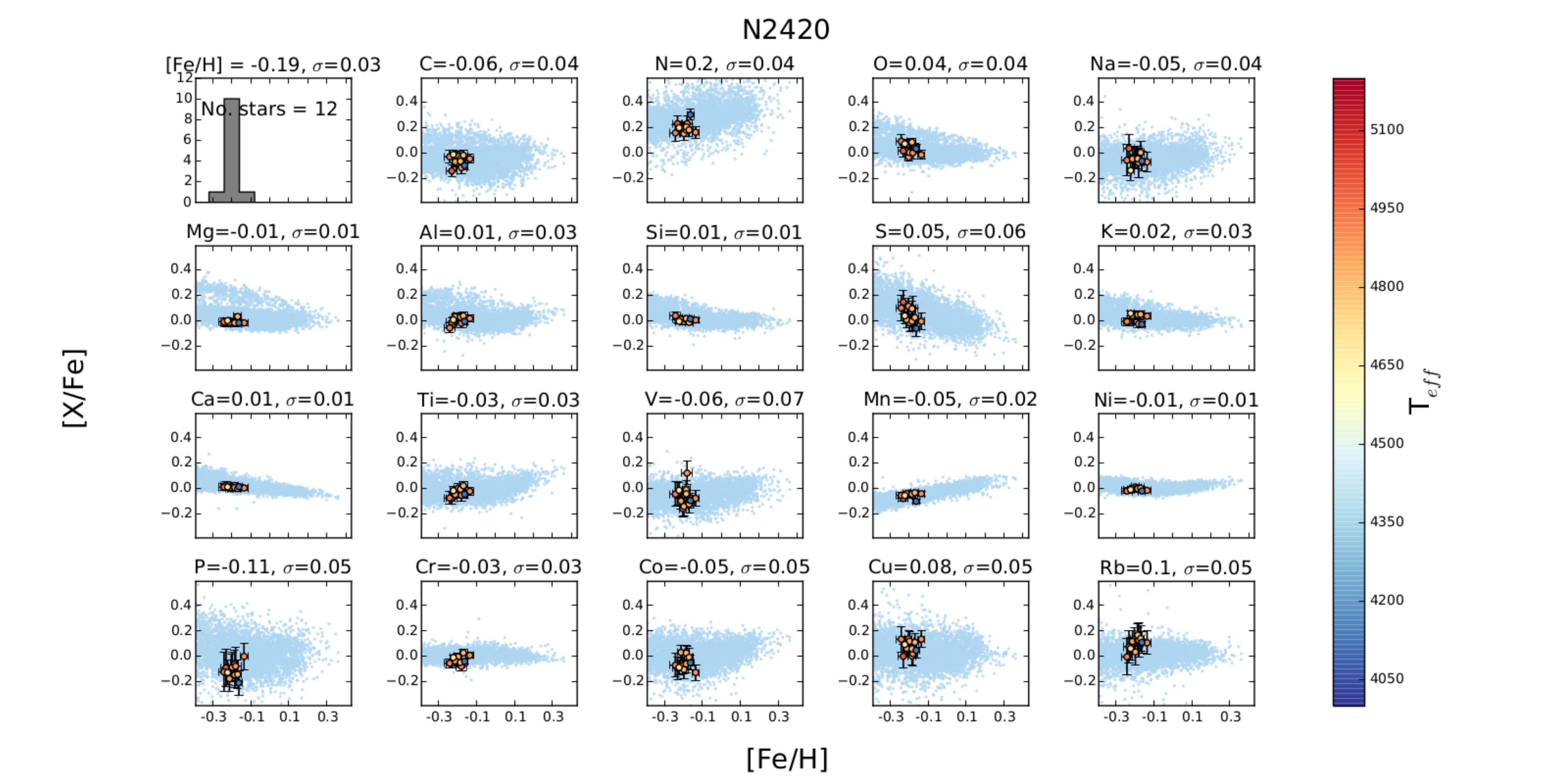}  
  \caption{ As per Figure \ref{fig:c1} but for NGC2420 stars with  a median SNR = 311. } 
\label{fig:c2}
\end{figure*}

\begin{figure*}
\centering
                     \includegraphics[scale=0.5]{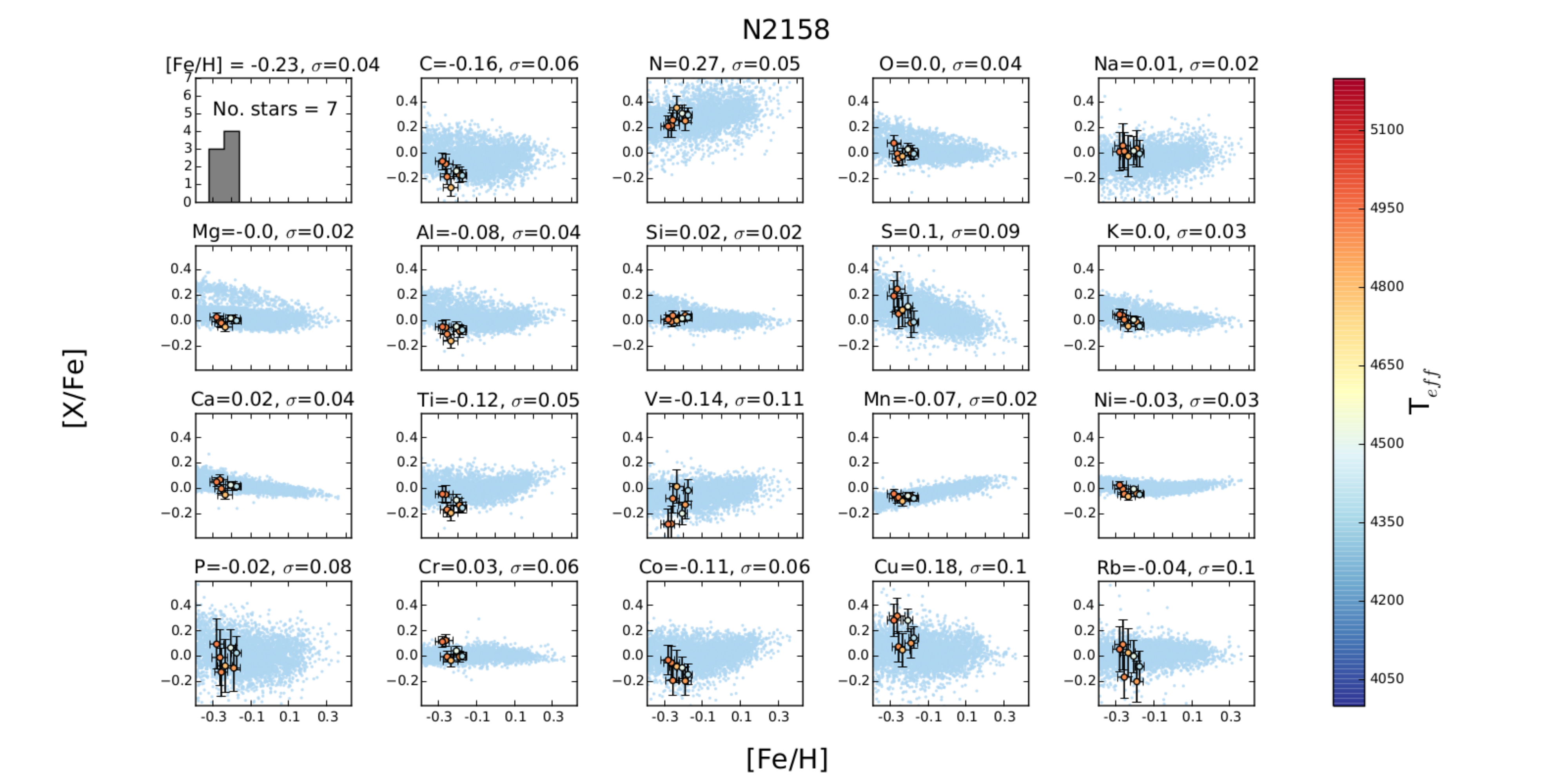}
  \caption{As per Figure \ref{fig:c1} but for NGC2158 stars with a median SNR = 96.  } 
\label{fig:c3}
\end{figure*}

\begin{figure*}
\centering
                 \includegraphics[scale=0.5]{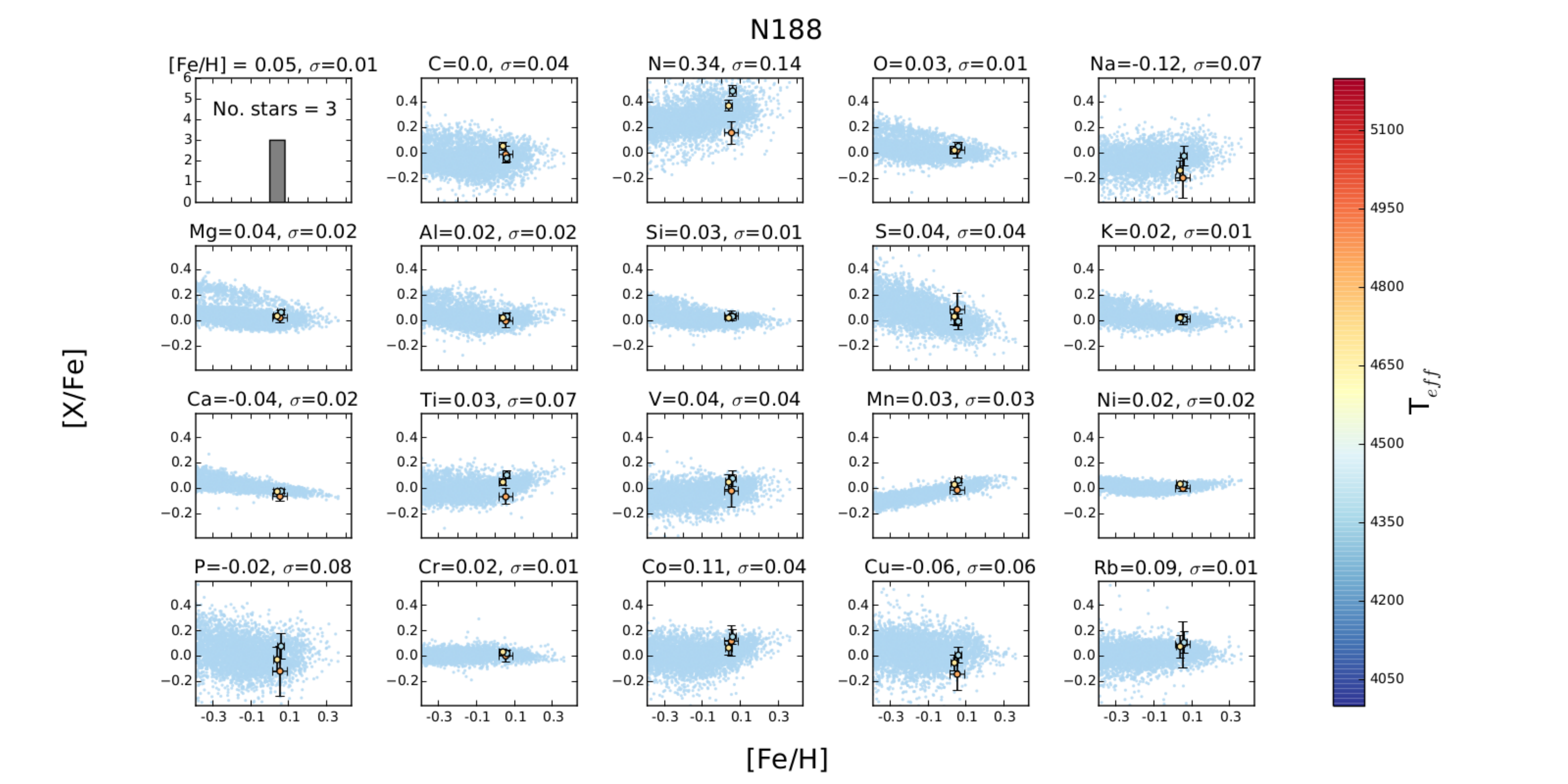}
  \caption{As per Figure \ref{fig:c1} but for NGC188 stars with a median SNR = 462. }
\label{fig:c4}
\end{figure*}

\begin{figure*}
\centering
    \includegraphics[scale=0.5]{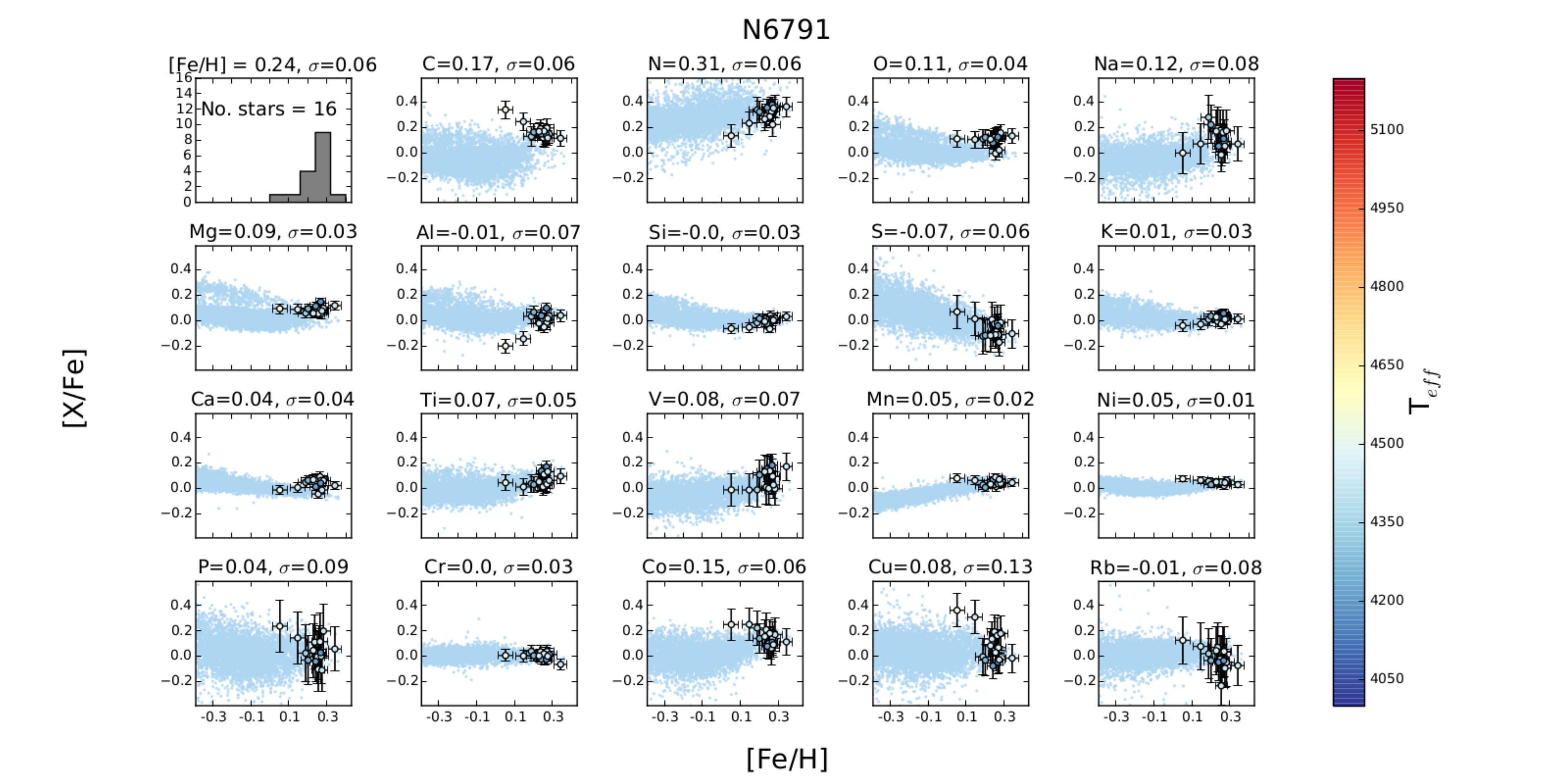}
  \caption{As per Figure \ref{fig:c1} but for NGC6791 stars with a median SNR = 159. } 
\label{fig:c6}
\end{figure*}

\begin{figure*}
\centering
        \includegraphics[scale=0.5]{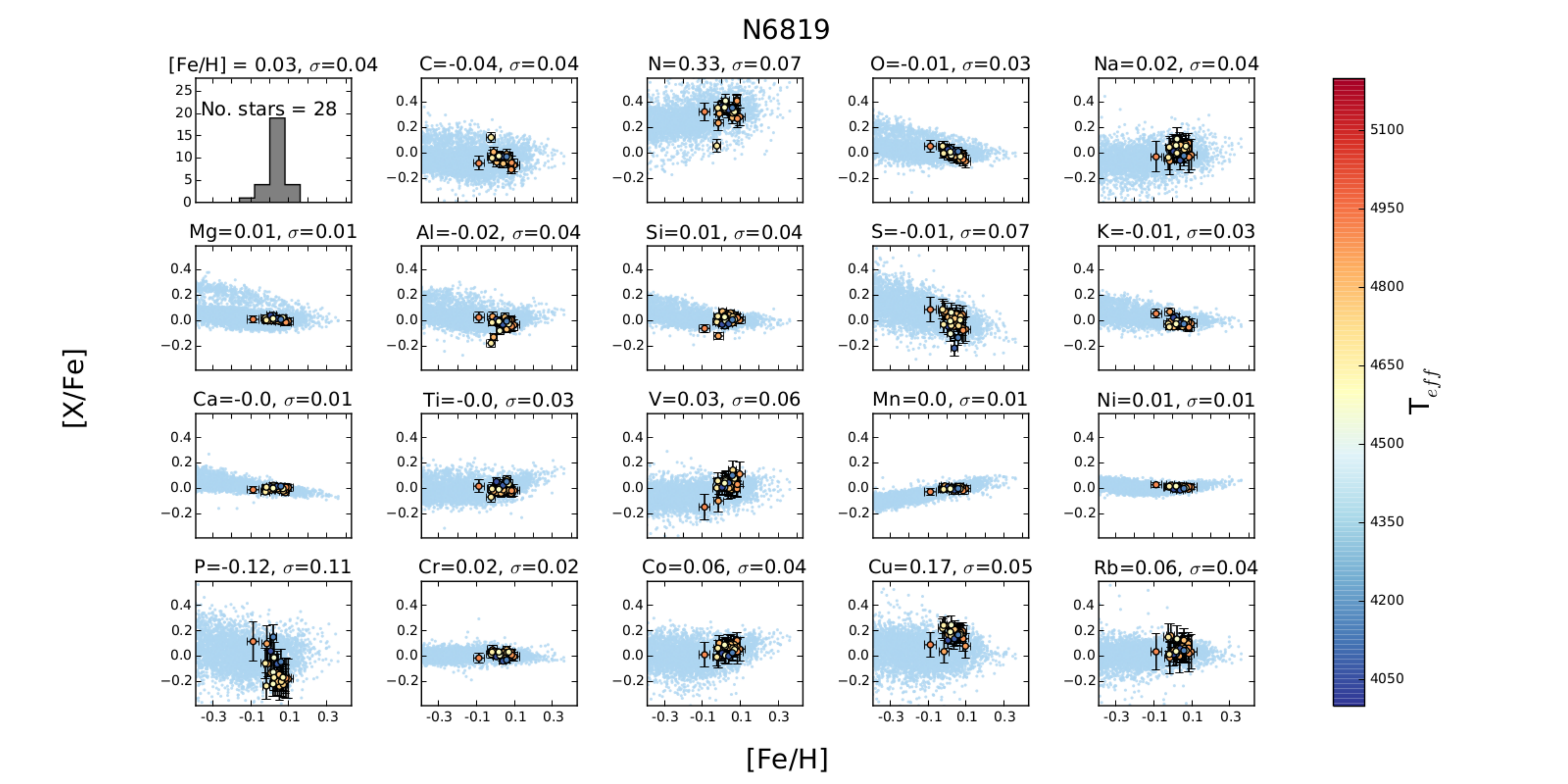}
  \caption{As per Figure \ref{fig:c1} but for NGC6819 stars with a median SNR = 303. } 
\label{fig:c7}
\end{figure*}

\begin{figure*}
\centering
  \includegraphics[scale=0.5]{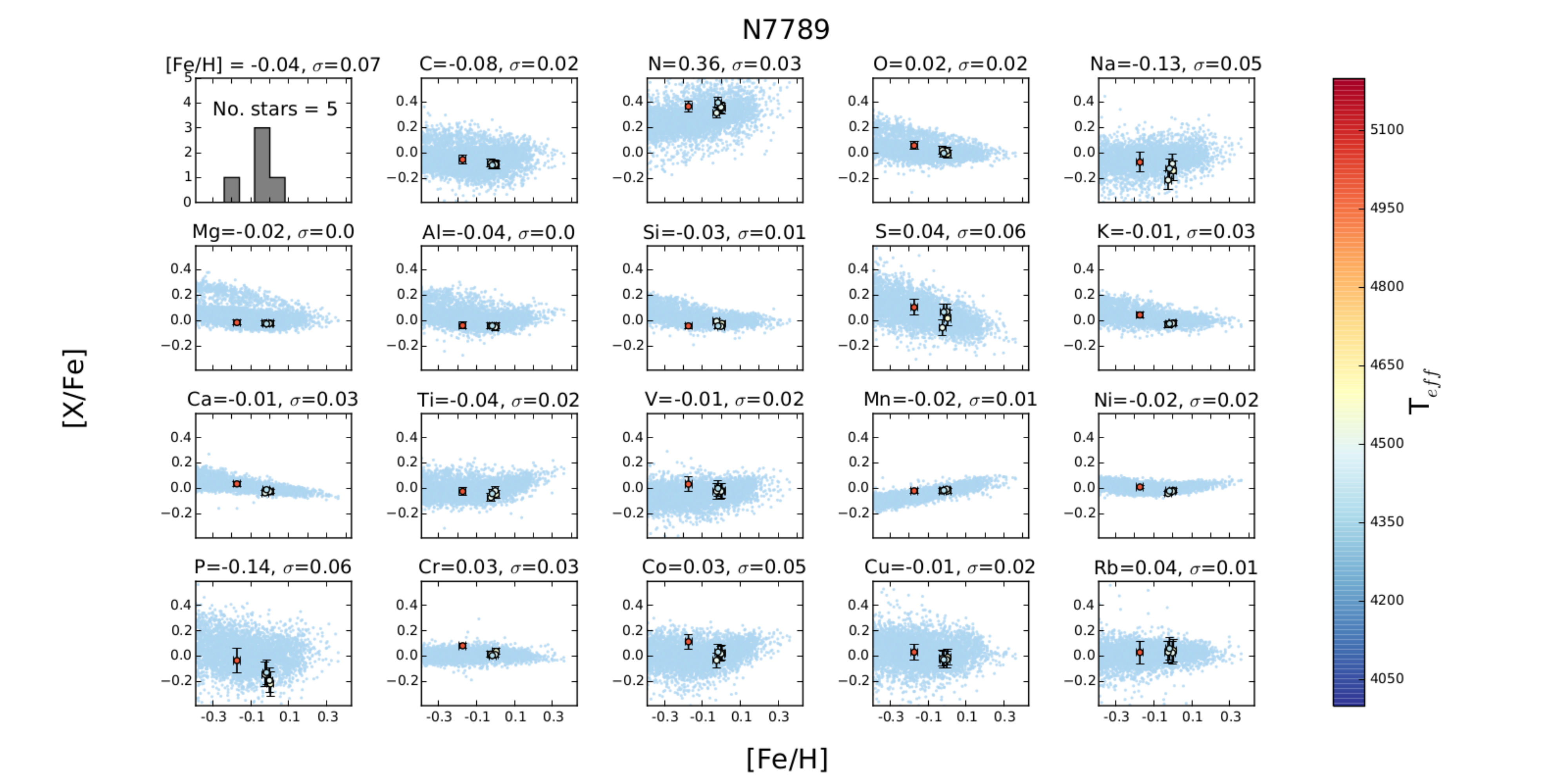}
  \caption{As per Figure \ref{fig:c1} but for NGC7789 stars with a median SNR = 657. } 
\label{fig:c8}
\end{figure*}

Figures \ref{fig:c1} to \ref{fig:c8} show the cluster individual abundances with their corresponding uncertainties. The points are colored by  \teff\ to check for any systematic abundance trends.  The typical 1-$\sigma$ standard deviation measurements for the elements around their mean values range from 0.01 -- 0.1.  The cyan points in the background are the labels for all cluster
stars used in the training of \tc . The DR13 input abundance labels that we adopt for training are already corrected for systematic variations with temperature \citep{Holtzman2015}, except for C and N, which are known to have astrophysical variations along the giant branch (and then additionally corrected as part of our procedure for LSF variation as described in Section 4). For the open clusters, we note that the coolest stars do seem to have the highest measurements of [N/Fe], within a cluster. 

We report the measured abundance mean and variance for each of our 90 open cluster stars in Table 1,
along with the corresponding \aspcap\ results (where available) and 2MASS IDs. Our results compare very well to the \aspcap\ results, but using \tc, we obtain substantively higher precision and so can report measurement uncertainties that are 20\% - 50\% lower than \aspcap, in most cases. Here we are concerned with precision and not really with accuracy; there is a discussion in \citet{Holtzman2015} regarding the scale of the open clusters with respect to the literature. Overall, we note there is a large variation in the reported individual element measurements from high resolution spectroscopy (e.g. see Table 3 of \citet{Souto2016} for a literature comparison of one of the open clusters, NGC2420). 

\floattable
\begin{deluxetable}{lccccc}
\tablecaption{ Open Cluster Properties \label{t:tab1}}
\tablecolumns{6}
\tablenum{1}
\tablewidth{0pt}
\tablehead{
\colhead{Cluster} & \colhead{l} & \colhead{b}  & \colhead{Distance (kpc)} & \colhead{$E(B-V)$} & \colhead{$log (Age)$} \\
}
\startdata
NGC7789 &     115.532 & -5.385 & 2.34 & 0.22 & 9.235 \\
NGC6819 &      73.978 &  8.481 & 2.36 & 0.24 & 9.174 \\
NGC6791 &      69.959 & 10.904 & 4.10 & 0.12 & 9.643 \\
NGC2420 &     198.107 & 19.634 & 3.09 & 0.03 & 9.048 \\
NGC2158 &     186.634 &  1.781 & 5.07 & 0.36 & 9.023 \\
NGC188  &     122.843 & 22.384 & 2.05 & 0.08 & 9.632 \\
M67(NGC2682)& 215.696 & 31.896 & 0.91 & 0.06 & 9.409 \\
\enddata
\tablecomments{The cluster data are from the WEBDA Open Cluster Database www.univie.ac.at/webda/}
\end{deluxetable}

\begin{table*}
\centering
\tablenum{2}
\caption{Measured mean abundances from \tc\ and from \aspcap\ for the cluster stars. All elements are with respect to Fe, [X/Fe] except Fe which is [Fe/H].}
\tiny
\begin{tabular}{ | p{0.04\textwidth} | c| c | c | c | c | c | c | c | }
\hline
\multicolumn{1}{|c|}{Element}& \multicolumn{2}{|c|}{NGC7789 (5 stars)} & \multicolumn{2}{|c|}{NGC6819 (28 stars)}  & \multicolumn{2}{|c|}{NGC6791 (16 stars)} & \multicolumn{2}{|c|}{ NGC188 (3 stars)} \\
\hline
 &  ASPCAP & The Cannon & APSCAP & The Cannon & ASPCAP & The Cannon & ASPCAP & The Cannon \\
 \hline
Fe &  -0.05 $\pm$ 0.05  & -0.04 $\pm$ 0.07 &  0.04 $\pm$ 0.04   & 0.03 $\pm$ 0.04  &  0.24 $\pm$ 0.06  & 0.21 $\pm$ 0.08 &   0.05 $\pm$ 0.02 & 0.05 $\pm$ 0.01   \\
C & -0.08 $\pm$ 0.07  & -0.08 $\pm$ 0.02' & -0.05 $\pm$ 0.09 & -0.04 $\pm$ 0.04 & 0.18 $\pm$ 0.1 &0.17 $\pm$ 0.06 & 0.0 $\pm$ 0.05 & 0.0 $\pm$ 0.04 \\
N &  0.37 $\pm$ 0.07 & 0.36 $\pm$ 0.03 & 0.33 $\pm$ 0.08 & 0.33 $\pm$ 0.07 & 0.32 $\pm$ 0.07 & 0.31 $\pm$ 0.076&  0.36 $\pm$ 0.17 & 0.34 $\pm$ 0.14 \\
O & 0.01 $\pm$ 0.08 & 0.02 $\pm$ 0.02 &  0.01 $\pm$ 0.06  & -0.01 $\pm$ 0.03 & 0.1 $\pm$ 0.07 & 0.11 $\pm$ 0.04 & 0.07 $\pm$ 0.03  & 0.03 $\pm$ 0.01 \\
Na & -0.03 $\pm$ 1.3 & -0.13 $\pm$ 0.05 &  0.01 $\pm$ 0.06  & 0.02 $\pm$ 0.04 & 0.09 $\pm$ 0.08 & 0.12 $\pm$ 0.08 &  0.08 $\pm$ 0.02 & -0.12 $\pm$ 0.07 \\
Mg  & -0.02 $\pm$ 0.05 & -0.02 $\pm$ 0.0 &   -0.0 $\pm$ 0.04 & 0.01 $\pm$ 0.01 & 0.1 $\pm$ 0.06 & 0.09 $\pm$ 0.03 &  0.03 $\pm$ 0.05 & 0.04 $\pm$ 0.02 \\
Al  & -0.07 $\pm$ 0.06 & -0.04 $\pm$ 0.0  & -0.02 $\pm$ 0.05 & -0.02 $\pm$ 0.04- & 0.1 $\pm$ 0.11  & -0.01 $\pm$ 0.07 & 0.03 $\pm$ 0.03  & 0.02 $\pm$ 0.02 \\
Si & -0.01 $\pm$ 0.08 & -0.03 $\pm$ 0.01 &  0.02 $\pm$ 0.05 & 0.01 $\pm$ 0.04 & 0.14 $\pm$ 0.06  & 0.0 $\pm$ 0.03 & 0.04 $\pm$ 0.01  & 0.03 $\pm$ 0.01  \\
S & -0.02 $\pm$ 0.03 &  0.04 $\pm$ 0.06 &  0.01 $\pm$ 0.05 & -0.01 $\pm$ 0.07  & 0.02 $\pm$ 0.06 & -0.07 $\pm$ 0.06 & 0.0 $\pm$ 0.05   &  0.04 $\pm$ 0.04 \\
K & 0.11$\pm$ 0.1 & -0.01 $\pm$ 0.03 & -0.01 $\pm$ 0.1 & -0.01 $\pm$ 0.03 &  0.09 $\pm$ 0.15  & 0.01 $\pm$ 0.03 & -0.04 $\pm$ 0.05  & 0.02 $\pm$ 0.01  \\
Ca & -0.01$\pm$ 0.08 & -0.01 $\pm$ 0.03 & -0.01 $\pm$ 0.05  & 0.0 $\pm$ 0.01 & 0.02 $\pm$ 0.08 & 0.04 $\pm$ 0.04 &  -0.03 $\pm$ 0.07 & -0.04 $\pm$ 0.02 \\
Ti & -0.01$\pm$ 0.06 & -0.04 $\pm$ 0.02 & 0.01 $\pm$ 0.04  & 0.0 $\pm$ 0.03 & 0.02 $\pm$ 0.09 & 0.07 $\pm$ 0.05 &  -0.03 $\pm$ 0.03 & 0.03 $\pm$ 0.07 \\
V & -0.01$\pm$ 0.1 &  -0.01 $\pm$ 0.02 & 0.01 $\pm$ 0.06 & 0.03 $\pm$ 0.06 & 0.07 $\pm$ 0.14 & 0.08 $\pm$ 0.07 & 0.0 $\pm$ 0.05  & 0.04 $\pm$ 0.04 \\
Mn & -0.02 $\pm$ 0.06 & -0.02 $\pm$ 0.01  &  0.0 $\pm$ 0.04 & 0.0 $\pm$ 0.01 & 0.02 $\pm$ 0.09 & 0.05 $\pm$ 0.02 & 0.01 $\pm$ 0.06  & 0.03 $\pm$ 0.03 \\
Ni & -0.02 $\pm$ 0.06 & -0.02 $\pm$ 0.02  &  0.01 $\pm$ 0.04 & 0.01 $\pm$ 0.01 & 0.03 $\pm$ 0.07 & 0.05 $\pm$ 0.01 & 0.03 $\pm$ 0.03  & 0.02 $\pm$ 0.02 \\
P & -0.06 $\pm$ 0.07 &  -0.14 $\pm$ 0.06  & -0.05 $\pm$ 0.15 & -0.12 $\pm$ 0.11 & 0.06 $\pm$ 0.11 & 0.06 $\pm$ 0.09 & 0.08 $\pm$ 0.03  & -0.02 $\pm$ 0.08 \\
Cr & 0.02 $\pm$ 0.08 &  0.03 $\pm$ 0.03 &  0.01 $\pm$ 0.05 & 0.02 $\pm$ 0.02 & -0.08 $\pm$ 0.0 &0.02 $\pm$ 0.03 & -0.04 $\pm$ 0.08  &  0.02 $\pm$ 0.01 \\
Co & -0.01$\pm$ 0.11 & 0.03 $\pm$ 0.05 &  0.04 $\pm$ 0.07 & 0.06 $\pm$ 0.04 & 0.17 $\pm$ 0.07 & 0.15 $\pm$ 0.06 &  0.15 $\pm$ 0.07 &  0.11 $\pm$ 0.04 \\
Cu &  0.0 $\pm$ 0.1 & -0.01 $\pm$ 0.02 & 0.1 $\pm$ 0.08 & 0.17 $\pm$ 0.05 & -- $\pm$ 0.0 & 0.08 $\pm$ 0.13 &  -0.03 $\pm$ 0.11 & -0.06 $\pm$ 0.06 \\
Rb & 0.05 $\pm$ 0.05 & 0.04 $\pm$ 0.01  &  0.01 $\pm$ 0.07 & 0.06 $\pm$ 0.04 & 0.01 $\pm$ 0.11 & -0.01 $\pm$ 0.08 & 0.06 $\pm$ 0.02  & 0.09 $\pm$ 0.01 \\
 \hline
\multicolumn{1}{|c|}{Element}& \multicolumn{2}{|c|}{ NGC2420 (12 stars)} & \multicolumn{2}{|c|}{NGC2158 (7 stars)}  & \multicolumn{2}{|c|}{M67 (19 stars)} & \multicolumn{2}{|c|}{}  \\
\hline
 &  ASPCAP & The Cannon & APSCAP & The Cannon & ASPCAP & The Cannon &  &   \\
 \hline
Fe & -0.18 $\pm$ 0.02 & -0.19 $\pm$ 0.03 &  -0.19 $\pm$ 0.04  & -0.23 $\pm$ 0.04 &  0.0 $\pm$ 0.04 &0 $\pm$ 0.03 &  &  \\
C & -0.06 $\pm$ 0.04  &-0.06 $\pm$ 0.04 & -0.11 $\pm$ 0.13 & -0.16 $\pm$ 0.06 & -0.11 $\pm$ 0.08 &  -0.09 $\pm$ 0.05 &  &  \\
N &  0.21 $\pm$ 0.05 & 0.2 $\pm$ 0.04 & 0.25 $\pm$ 0.08  & 0.27 $\pm$ 0.05 & 0.35 $\pm$ 0.09  & 0.33 $\pm$ 0.07  &  &  \\
O & 0.04 $\pm$ 0.04 &0.04 $\pm$ 0.04 &  -0.05 $\pm$ 0.12  & 0.0 $\pm$ 0.04 & -0.03 $\pm$ 0.06 & -0.03 $\pm$ 0.02 &  &  \\
Na & -0.01 $\pm$ 0.04 & -0.05 $\pm$ 0.04 & 0.03 $\pm$ 0.12   & 0.01 $\pm$ 0.02  & 0.0 $\pm$ 0.08  & -0.01 $\pm$ 0.03 &  &  \\
Mg  & -0.02 $\pm$ 0.03 & -0.01 $\pm$ 0.01 & 0.0 $\pm$ 0.06   &0.0 $\pm$ 0.02 & 0.0 $\pm$ 0.04 & 0.01 $\pm$ 0.02 &  &  \\
Al  & -0.02 $\pm$ 0.03 & 0.01 $\pm$ 0.03  & -0.03 $\pm$ 0.05 & -0.08 $\pm$ 0.04 & -0.04 $\pm$ 0.05  & -0.03 $\pm$ 0.02  &  &  \\
Si & -0.03 $\pm$ 0.03 & 0.01 $\pm$ 0.01 & -0.07 $\pm$ 0.06  & 0.02 $\pm$ 0.02 & -0.02 $\pm$ 0.03 & -0.02 $\pm$ 0.03 &  &  \\
S & 0.0 $\pm$ 0.02 &  0.05 $\pm$ 0.06 &  0.02 $\pm$ 0.02 & 0.1 $\pm$ 0.09 & -0.02 $\pm$ 0.04 &-0.04 $\pm$ 0.08 &  &  \\
K & 0.07 $\pm$ 0.07  &  0.02 $\pm$ 0.03  & 0.16 $\pm$ 0.13 & 0.0 $\pm$ 0.03 &  -0.01 $\pm$ 0.07   &-0.03 $\pm$ 0.01 &  &  \\
Ca & 0.02 $\pm$ 0.05  &  0.01 $\pm$ 0.01 & 0.03 $\pm$ 0.08  & 0.02 $\pm$ 0.04 & -0.02 $\pm$ 0.04 & -0.01 $\pm$ 0.02 &  &  \\
Ti & 0.02 $\pm$ 0.03  &  -0.03 $\pm$ 0.03  &  -0.01 $\pm$ 0.04  & -0.12 $\pm$ 0.05 & -0.02 $\pm$ 0.04 & -0.03 $\pm$ 0.04 &  &  \\
V & -0.05 $\pm$ 0.04 &  -0.06 $\pm$ 0.07 &  -0.1 $\pm$ 0.08  & -0.14 $\pm$ 0.11  &  -0.03 $\pm$ 0.08 & -0.01 $\pm$ 0.06  &  &  \\
Mn & -0.06 $\pm$ 0.03 & -0.05 $\pm$ 0.02  &   -0.07 $\pm$ 0.06 & -0.07 $\pm$ 0.02 & -0.02 $\pm$ 0.04 &-0.03 $\pm$ 0.01 &  &  \\
Ni & -0.02 $\pm$ 0.03 & -0.01 $\pm$ 0.01 &  -0.03 $\pm$ 0.06 & -0.03 $\pm$ 0.03 & 0.01 $\pm$ 0.04  & 0.01 $\pm$ 0.01 &  &  \\
P & -0.05 $\pm$ 0.07 &  -0.11 $\pm$ 0.05  &  -0.01 $\pm$ 0.18 & -0.02 $\pm$ 0.08 & -0.09 $\pm$ 0.07 & -0.07 $\pm$ 0.05  &  &  \\
Cr & -0.04 $\pm$ 0.06 &  -0.03 $\pm$ 0.03 &  -0.01 $\pm$ 0.17  & 0.03 $\pm$ 0.06 & -0.01 $\pm$ 0.06  &-0.01 $\pm$ 0.02  &  &  \\
Co & -0.12 $\pm$ 0.08 & -0.05 $\pm$ 0.05  & -0.07 $\pm$ 0.08 & -0.11 $\pm$ 0.06 & -0.02 $\pm$ 0.06 & 0.04 $\pm$ 0.06 &  &  \\
Cu & 0.03 $\pm$ 0.08  &  0.08 $\pm$ 0.05  & 0.04 $\pm$ 0.24 & 0.18 $\pm$ 0.1  & 0.02 $\pm$ 0.11  & 0.12 $\pm$ 0.05  &  &  \\
Rb & 0.06 $\pm$ 0.09 & 0.1 $\pm$ 0.05  & -0.07 $\pm$ 0.22  & -0.04 $\pm$ 0.1 & 0.03 $\pm$ 0.06  & 0.06 $\pm$ 0.04  &  &  \\
 \hline
\end{tabular}
\ref{tab2}
\end{table*}

\subsubsection{Re-scaling some of the Abundance Uncertainties}
\label{uncertainty_rescale}

The quadrature sum of the formal abundance uncertainty for each star and 
the cross-validation error (typically $10\times$~larger; see Figure \ref{fig:cross})
represents the most conservative uncertainty estimate for our measurements,
essentially an upper limit on the precision; if these uncertainties were too large, the inferred intrinsic dispersion would be too low. 

\begin{figure*}
\centering
  \includegraphics[scale=0.5]{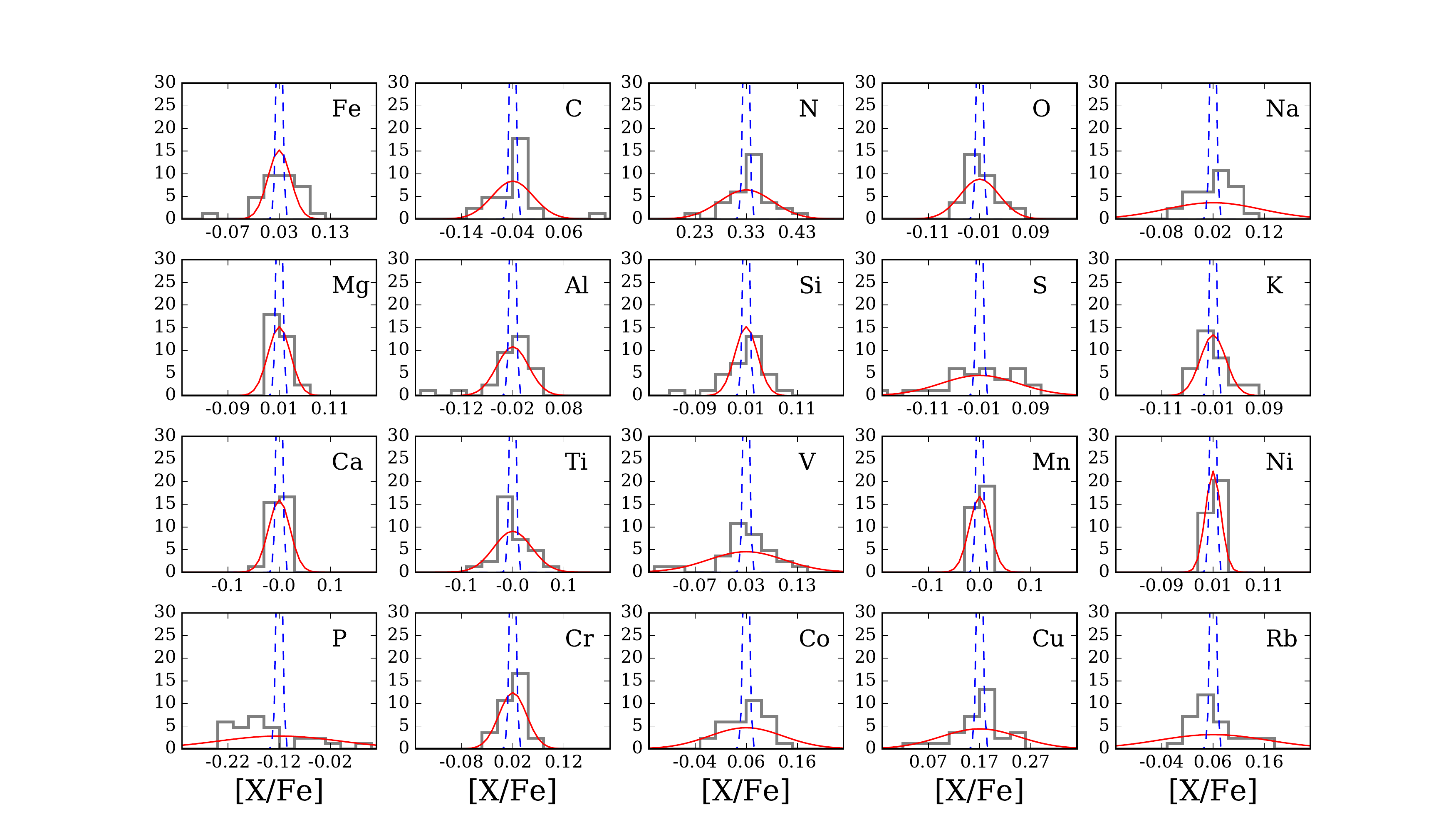}
   \caption{NGC6819 individual abundance distributions; all elements are with respect to Fe except for Fe, which is [Fe/H]. The y-axis shows the number of stars. The black histograms show the binned data itself, the red Gaussians show the cross-validation errors and the narrow blue histograms show the median formal errors for these stars, distributed around the mean abundance measurement for each  abundance measurement returned by \tc.  }
\label{fig:minmax_error}
\end{figure*}

We check this using the best sampled cluster,
NGC6819, with 28 stars, {\sl assuming} it is chemically homogeneous; the distribution of measured abundance estimates must not be ``narrower" than the abundance uncertainties. 
Figure \ref{fig:minmax_error} shows the abundance distribution for all elements for NGC6819. The data are shown in the binned histograms and the cross-validation errors are shown in the red Gaussians. The  median formal errors for the stars are shown in the narrow blue Gaussian distributions. Under the assumption that the cluster does not have any intrinsic dispersion in any element,  the cross-validation errors appear to be a relatively accurate representation of the measurement uncertainties. However, 
the uncertainties for some elements, for example, both [Na/Fe] and [V/Fe] look to be overestimated from this Figure, comparing the wide red Gaussian distributions to the  narrower width of the binned data. 

The $\chi^2$ distribution of pairs of stars within a cluster asks how likely the abundance measurements of a pair of stars would be if the true abundances were identical. We can therefore use the $\chi^2$ metric, for all pairs of stars within a cluster, to check whether our uncertainties are over- or under-estimates. The $\chi^2$ value is given by the following equation: 

\begin{equation}
\chi_{nn'}^2 =  \frac{[x_{ni} - x_{n'i}]^2}{\sigma_{ni}^2 + \sigma_{n'i}^2},
\end{equation}

where the indices $n$ and $n'$ denote the two stars, $i$ the elements, and $x_{ni}$ the measurements
with uncertainty $\sigma_{ni}$.

\begin{figure*}
\centering
  \includegraphics[scale=0.5]{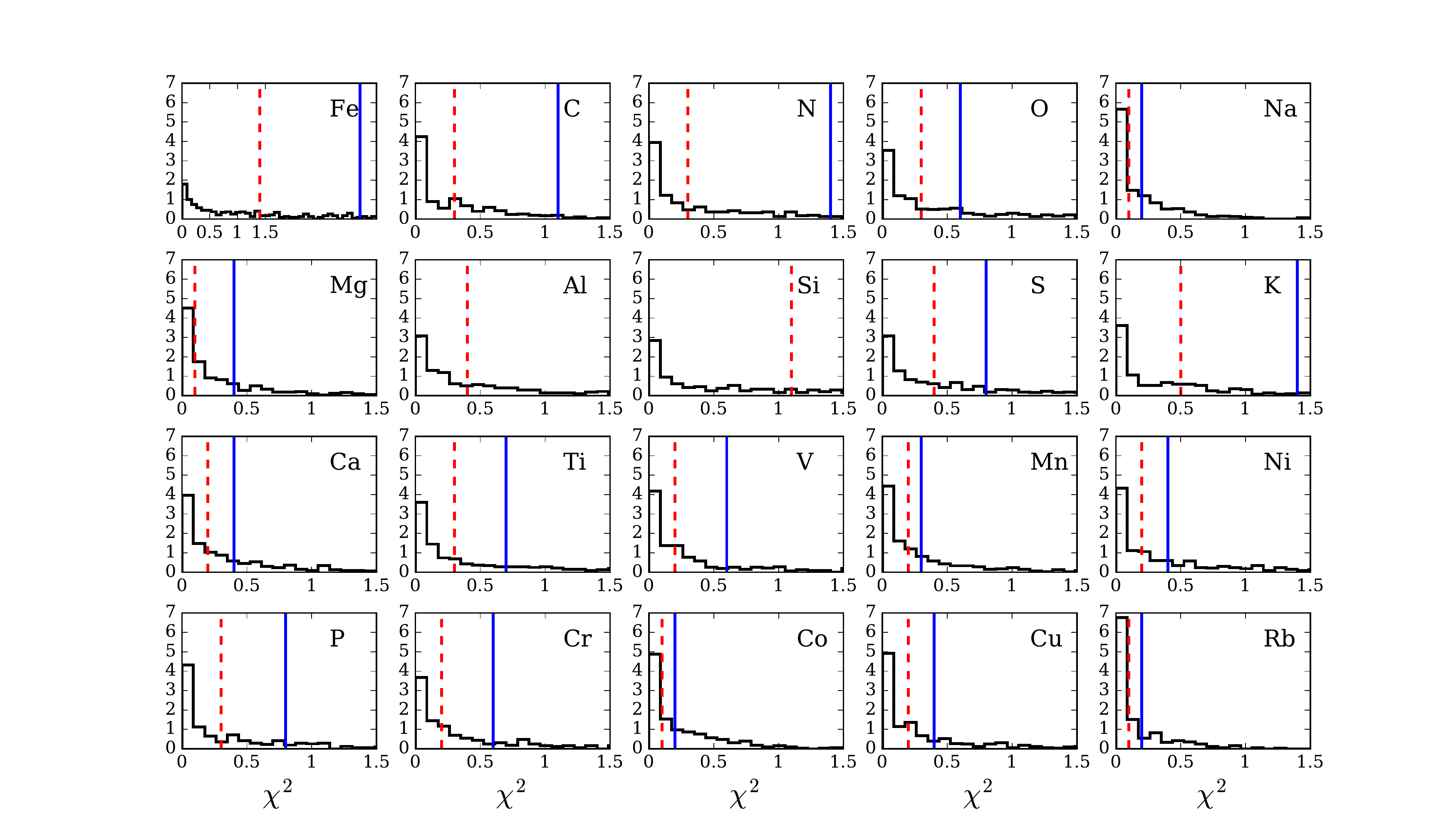}
   \caption{The $\chi^2$ distribution of all pairs of stars within the most populous cluster, NGC6819, which has 28 stars where the y-axis is the normalised number of stars.  All elements are measured with respect to Fe except for Fe, which is [Fe/H].  The mean values of the distributions are shown in the vertical blue lines and the median values are shown in the red dashed vertical lines. These mean and median values give the scaling values for the errors for each element, under the assumption of zero-dispersion for each abundance of the cluster, in which case the $\chi^2$ distribution would be expected to have a mean of 1 and a median of 0.45, for each element.   } 
 \label{fig:chi2_uncertainties}
\end{figure*}

Assuming the true abundances of the cluster stars were identical, we expect a 
distribution of $\chi_{nn'}^2$ that has a mean of $1$ and a median of $0.45$ 
for every element.  Figure \ref{fig:chi2_uncertainties} shows the  distribution of 
the $\chi_{nn'}^2$ values for all pairs of stars in our fiducial cluster, NGC6819, for each of our 20 elements. This Figure shows that the median and mean values of the distributions for a number of elements  have a mean $\chi^2< 1$ and a median $\chi^2< 0.45$, implying that the measurement uncertainties adopted so far must be overestimates. 
For those elements we decrease the uncertainties by the 
factors that make the mean or median of $\chi^2$ match the theoretical expectations.  

\subsection{Individual Abundance Dispersion within Clusters?}

Determining whether or not open clusters have a measurable intrinsic abundance spread is critical for the pursuit of chemical tagging and understanding the formation of these systems \citep[][]{Bovy2016, Liu2016}. With the carefully evaluated measurement uncertainties at hand, we are now in a position to determine these intrinsic spreads. At face value, the dispersion of abundance measurements among stars in a given cluster is comparable to the measurement uncertainties for the individual stars for each element. Qualitatively, this implies that the clusters are nearly homogeneous in their abundances. 
However, we need to determine formally for each element and for each cluster, what the {\sl intrinsic} dispersion is, given our data. 

Calculating the ``observed" abundance dispersion of a cluster can be done by simply fitting a Gaussian to the ensemble of abundance estimates \citep[e.g.][]{Ting2016}.  But here, we need the
``intrinsic" dispersions, which requires us to adopt a simple Gaussian model for the {\it intrinsic} abundance distribution $P(x_i)$ of the $i$-th element within a cluster, which 
accounts explicitly for the measurement uncertainties: \\

\begin{multline}
P({x^o_{i,n}} | \bar{x_i}, \sigma_i, \delta x_{i,n}) = \\
\prod_{n=1}^{N}
1/\sqrt{2 \pi (\delta x_{i,n}^2 + \sigma_i^2)}
\cdot\exp{\left ( - \frac{(\bar{x^o_{i,n}} - x_{i,n})^2}{2(\delta x_{i,n}^2 + \sigma_n^2)}\right )}
\end{multline}

Here the mean abundance of each element is $\bar{x}_i$ and the {\it intrinsic} (error corrected) abundance dispersion of each element within a cluster is $\sigma_i$. Our observational information is $\{x^o_{i,n},\delta x_{i,n}\}$. Of course, the intrinsic dispersion must be positive definite.
Then we are in a position to calculate the {\it pdf} for the intrinsic dispersion in each cluster. 

We also optimize the right hand side of equation (2) over a range of sigma-clipping values of the input data ($x_{i, n}$, $\delta x_{i, n}$) and take the maximal value. The typical optimal sigma-clipping is $\sigma_{clip}>$ 3. This means that only a few stars are excised by sigma clipping and for just a few of the elements. For clusters M67, NGC2420, NGC6791, NGC6819 and NGC7789, between 1-3 stars excised, for only a few elements; as few as 1 element (in NGC2420) and as many as only 6 elements (in NGC6819). 

We should explore the impact of the slightly different choices for the 
measurement uncertainties, discussed in Section \ref{uncertainty_rescale}: (i) where we scale this error by the median scaling value calculated for each element for the $\chi^2$ distribution shown in Figure \ref{fig:chi2_uncertainties}; (ii) as (i) but scaling by the mean; and (iii) leaving the
uncertainties $\delta x_{i,n}^2$ to be the quadrature sum of the formal uncertainty from \tc\  and the signal-to-noise dependent cross-validation uncertainty for that element. We find that  all three  error estimates give almost identical results and we show here the results for the median scaling. We use the median scaled errors for all the analysis that follows in the paper. 

\begin{figure*}
\centering
   \includegraphics[scale=0.4]{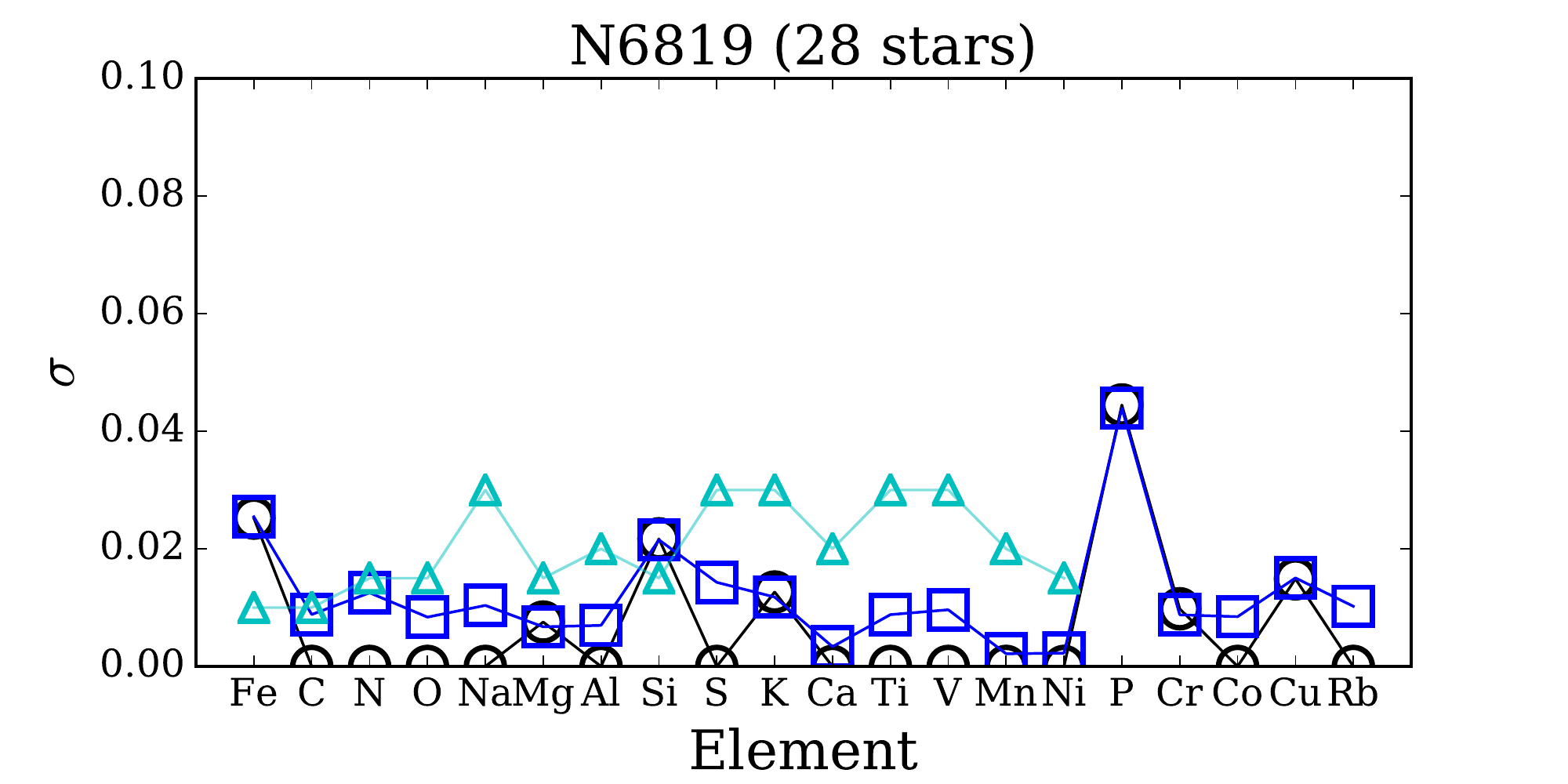}
 \includegraphics[scale=0.4]{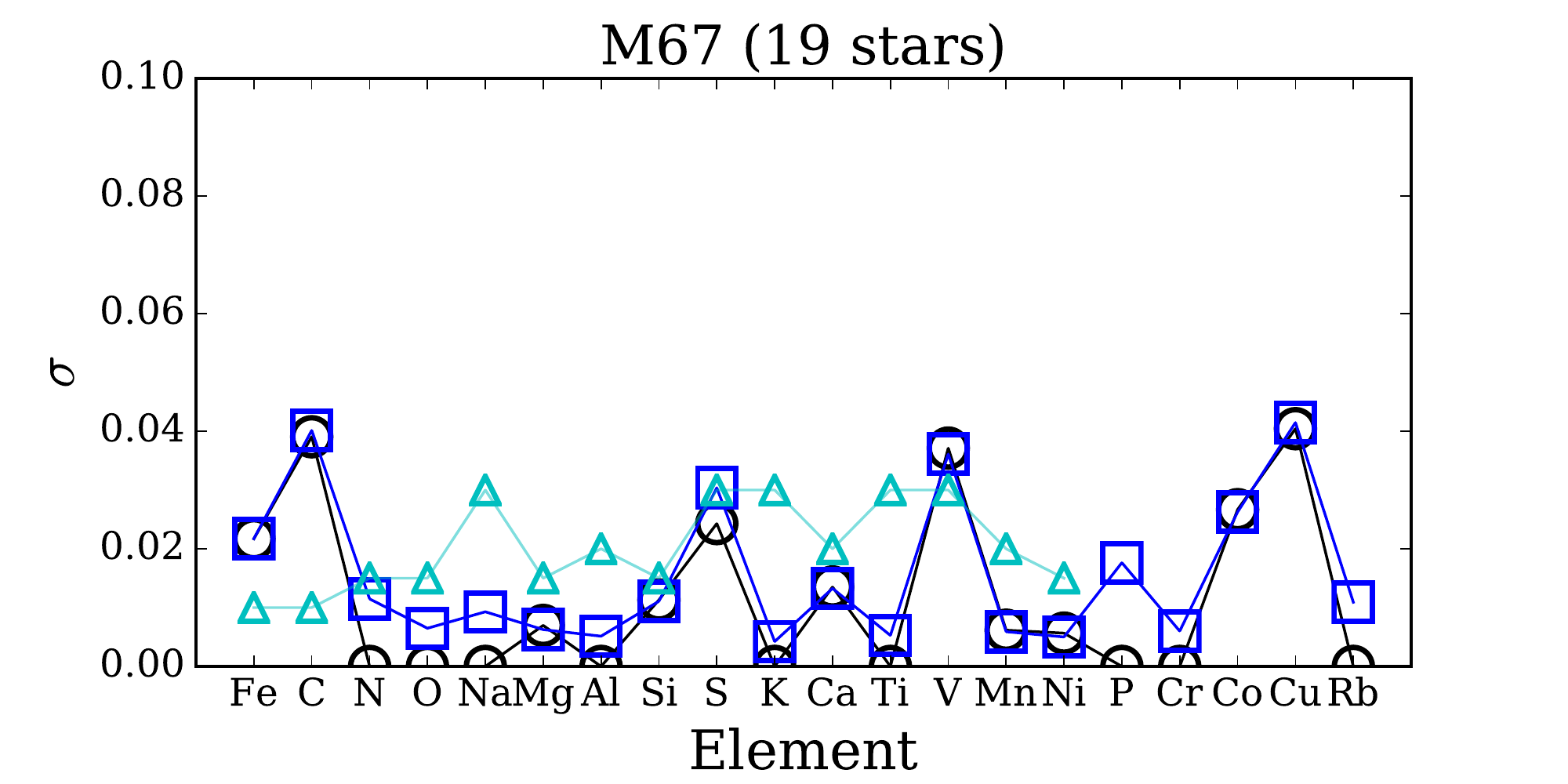}\\
   \includegraphics[scale=0.4]{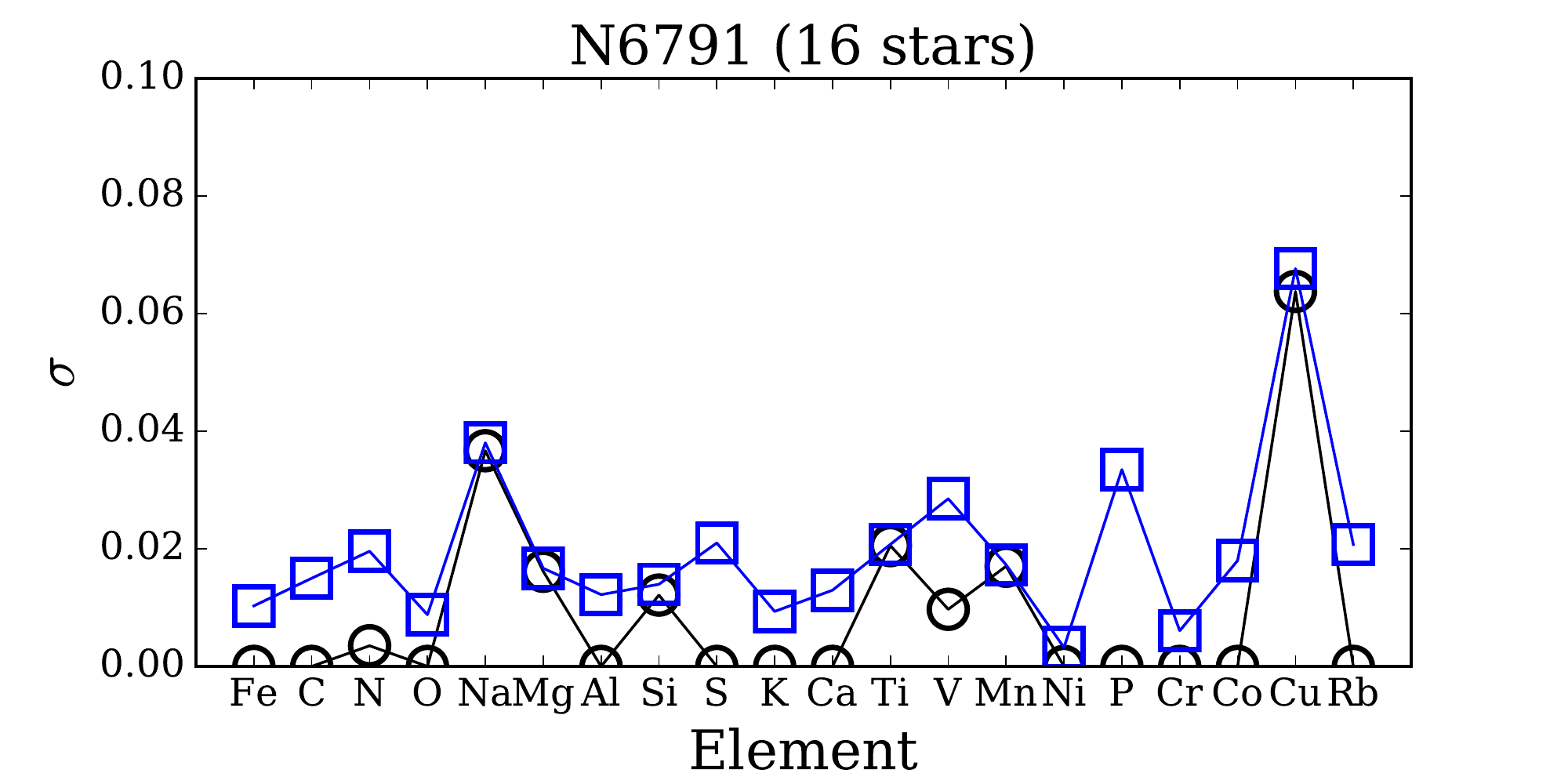}
     \includegraphics[scale=0.4]{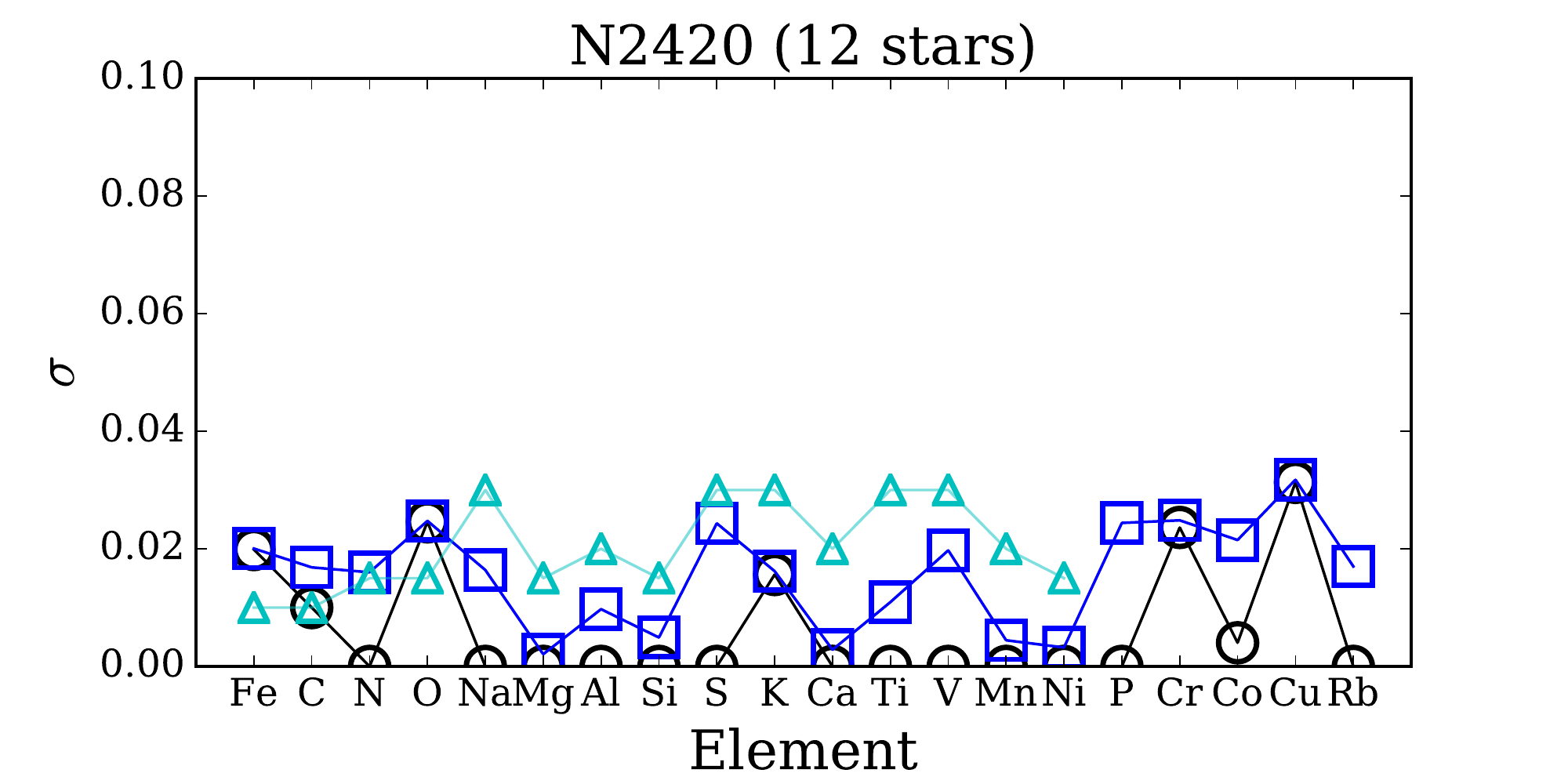}\\
  \includegraphics[scale=0.4]{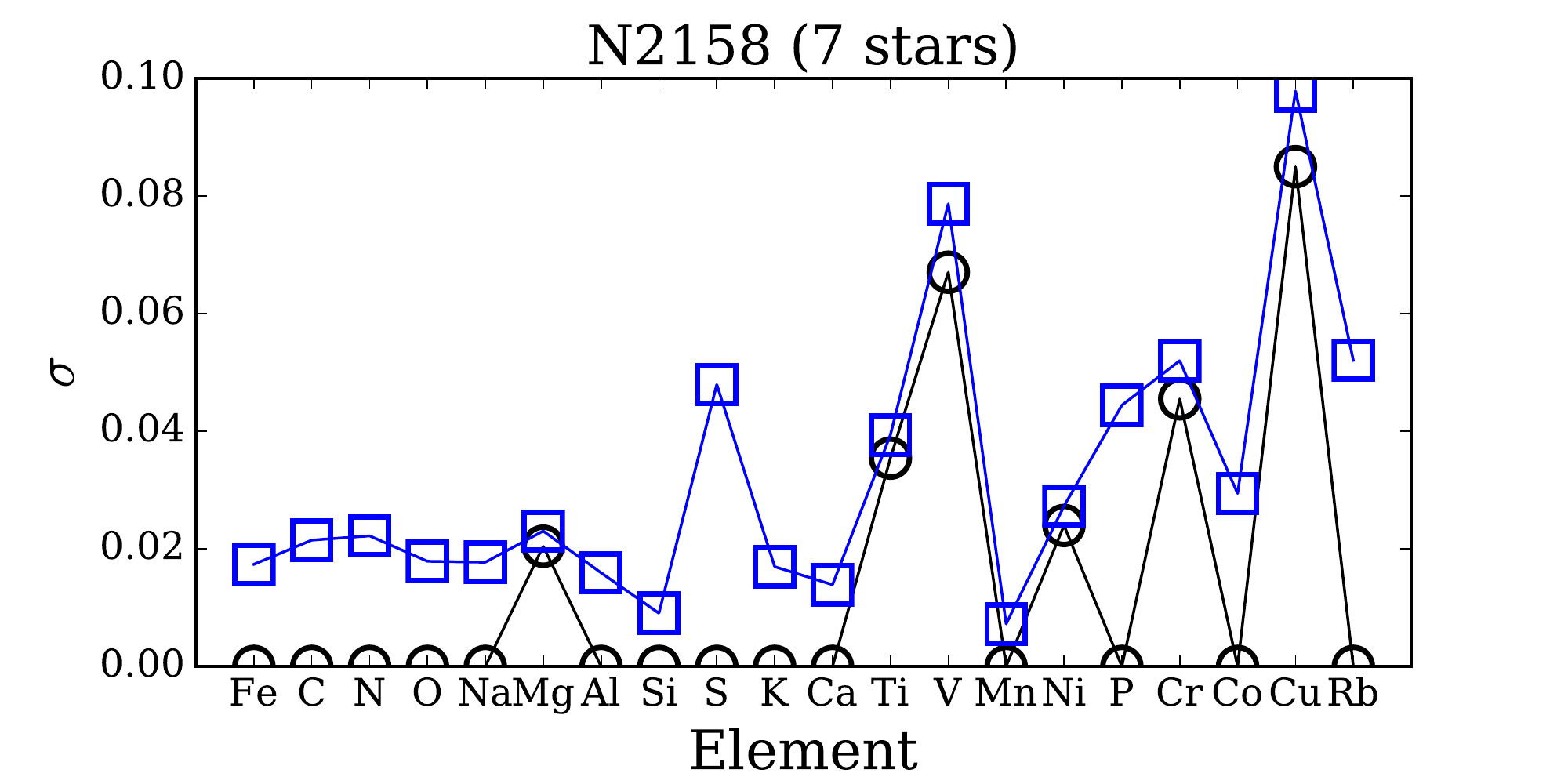}
   \includegraphics[scale=0.4]{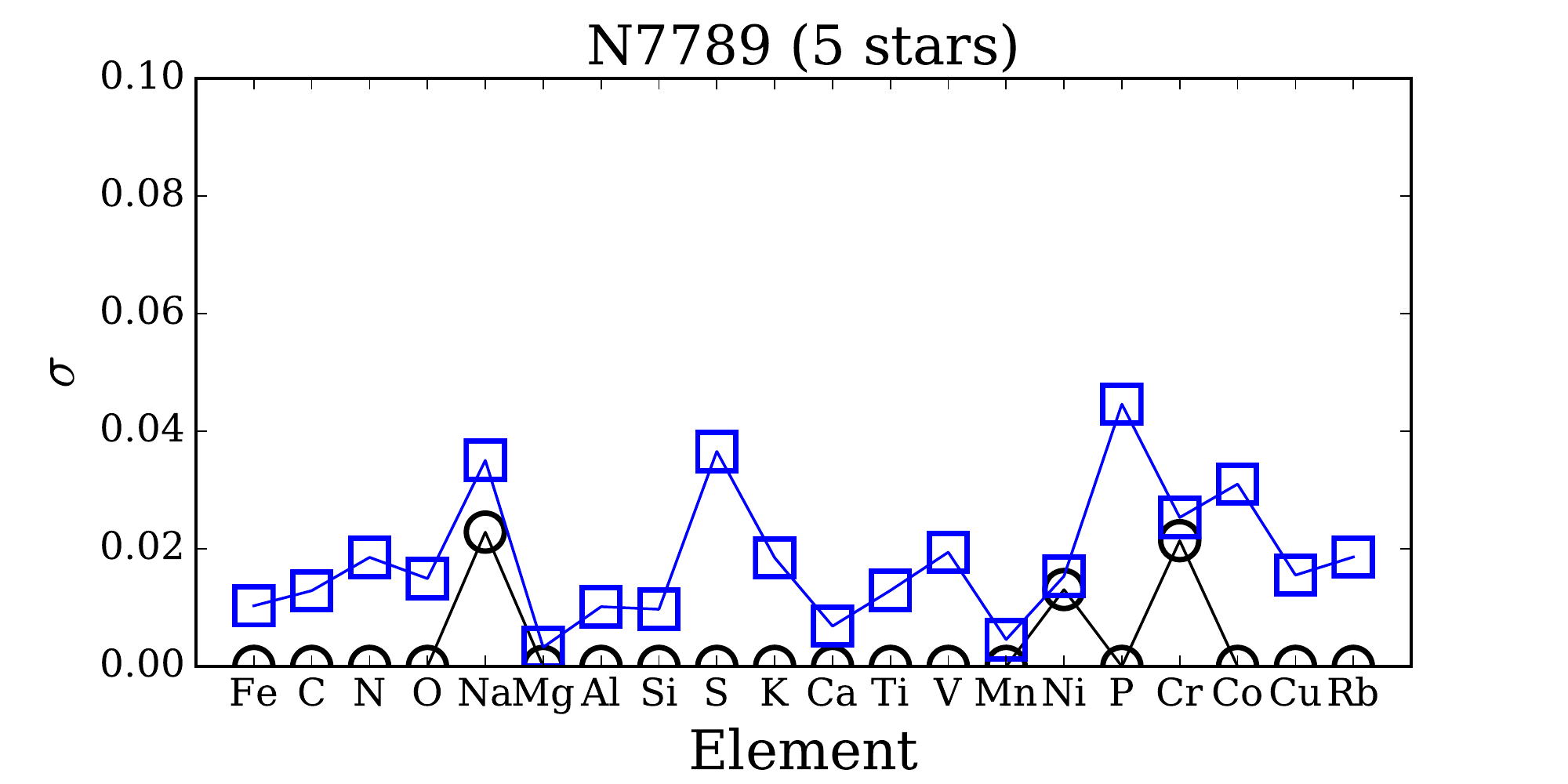} \\
     \includegraphics[scale=0.4]{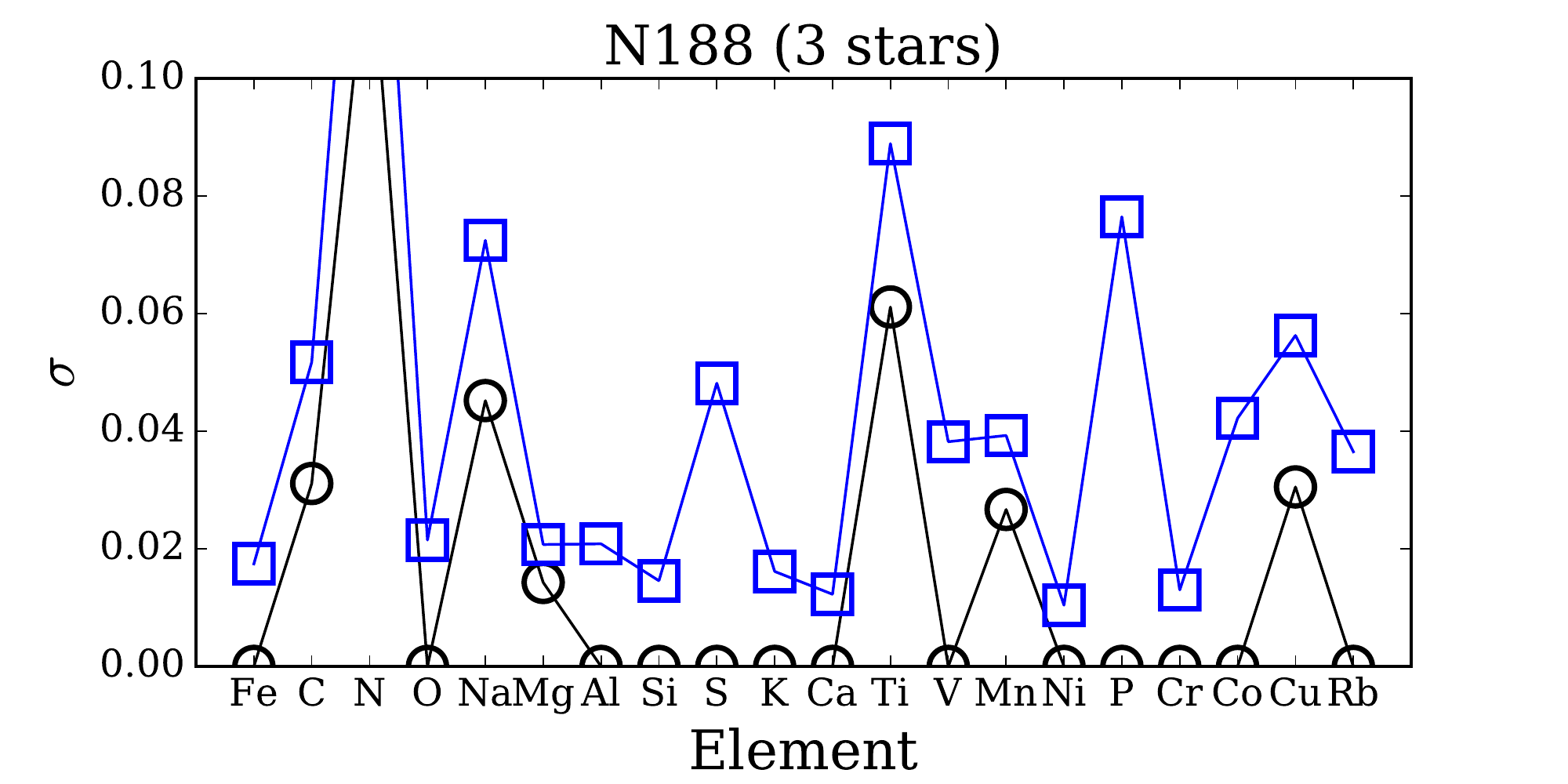}
  \caption{The intrinsic dispersion of the clusters showing  the most likely value in black circles and the 50th-percentile confidence interval value in blue squares, for all 20 elements and each cluster. The cyan triangles show the 68th-percentile confidence interval determined by \citet{Bovy2016}.  The number of stars are marked in each panel and clusters show a most likely intrinsic dispersion typically less than $<$ 0.02-0.03 dex. We have examined the impact of overestimated errors and the results we obtain are very similar for errors that are 10-20 percent lower, than our determined errors.  }
\label{fig:intrinsic}
\end{figure*}

Basically, we find throughout that the intrinsic abundance mean is very close to the mean of the measurements, 
as the abundance uncertainties among different stars within a cluster are similar. As the {\it pdfs} for the intrinsic abundance dispersions are by construction restricted to non-negative $\sigma_i$, we characterize them by two numbers; the most likely dispersion and the median of the {\it pdf}. This is shown in  Figure \ref{fig:intrinsic}: in most clusters and for
most elements the  intrinsic abundance dispersions are consistent with zero, and likely to be nearly zero.
The medians of the intrinsic dispersion \textit{pdfs} 
(Figure \ref{fig:intrinsic}) show that for all clusters with $>$ 10 stars (enabling robust dispersion estimates) the intrinsic dispersions are typically $<$ 0.02 dex, and $<$ 0.04 dex for almost all of the 20 elements. There are only a few stars with spectra in NGC2158, NGC 7789, and NGC188 and the measurements for  NGC188 in particular are not robust on the basis of small sample size. NGC188 has three stars and one is a clear outlier in some elements which drives the large intrinsic dispersion in this cluster. 

Our analysis points towards a small but finite intrinsic dispersion in [Fe/H] for two clusters with robust measurements (NGC6819 and M67), at the level of $\sigma$ = 0.02 dex;
this is true even if the abundance uncertainties were overestimated by 20\%. The  calculated probability of a zero dispersion value compared to the most likely value is only 
$p_{pdf}(\sigma_{[Fe/H]=0})/p_{pdf}(\sigma_{[Fe/H]_{most likely}})$ = 0.07.  
In general, these intrinsic abundance dispersion limits (or measurements) for clusters are roughly consistent with the limits placed on the elements by \citet{Bovy2016}. But here they are based on direct abundance determinations.


\subsection{Overall Abundance Similarity of Stars within Clusters and in the Field}

Based on these data we can try to set some expectation for how clearly abundances can tell us when stars in the \apogee\ sample are, or are  \textit{not}, born together in the same cluster, or molecular cloud core. We do this by examining the (dis)-similarity of stars within clusters and in the field. This is relevant for assessing the much larger sample of \apogee\ data and chemical dimensionality and diversity of the Milky Way disk. We note that our sample of open clusters does not span the full range of properties seen in the Galactic disk (i.e. their distribution in \feh,  age and Galactocentric radius). However, these clusters do have an abundance distribution that broadly resembles that of the \apogee\ field (disk) red clump sample. 

\begin{figure*}
\centering
\includegraphics[scale=0.45]{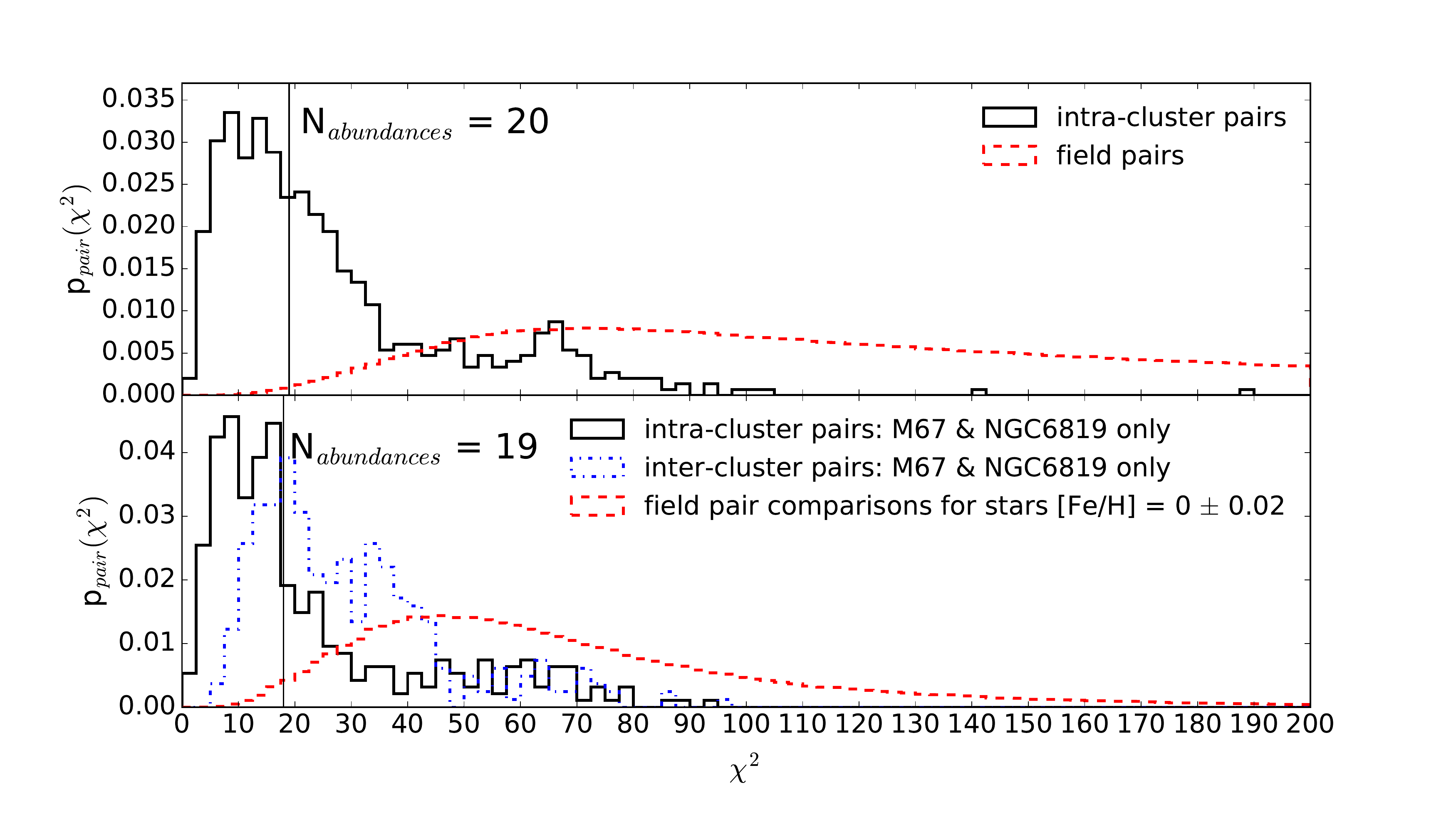} 
  \caption{At top: The $\chi^2$ distribution of abundance differences, $p_{pair}(\chi^2)$, for pairs with similar \logg\ and \teff\ values (see Section 4.3): the black histogram shows
  $p^{intra}_{pair}(\chi^2)$ for all intra-cluster pairs (600 pairs); the red dashed histogram shows the analogous distribution $p^{field}_{pair}(\chi^2)$ for all field pairs (1018581 pairs).
  The intra-cluster and field pair distributions are clearly very different,  but they are not disjoint: 0.3\% of field pairs have $\chi^2$ differences  as small as the median $\chi^2$ among intra-cluster pairs: most of these field pairs are presumably not birth siblings, but {\it doppelganger}. The bottom panel shows the analogous distributions $p_{pair}(\chi^2)$ to Fig. \ref{fig:within}, but restricted to pairs of the same [Fe/H], and the $\chi2$ arises from the sum over the 19 [X/Fe] estimates; the distribution is again restricted to pairs with similar \logg\ and \teff\ values (see text).
The [Fe/H] is set by the (near-identical) metallicity of the two clusters M67 and NGC6819. The $\chi^2$ distribution of the 377 intra-cluster pairs is shown as the black histogram. 
The distribution $p^{field}_{pair}(\chi^2)$ from 1054 random field stars with [Fe/H] = 0$\pm$0.02 and similar temperatures and gravities (301587 pairs) are shown in the red dashed histogram. The inter-cluster distribution for M67 and NGC6819 (of 327 pairs) is shown for comparison in the blue dash-dot histogram. Even if we select stars of the same [Fe/H], the other element abundances can still discriminate between intra-cluster and field pairs in the many instances: the peaks of the distributions are clearly different with the highest similarity among stars within a cluster and the largest dissimilarity among field stars. However, once again these distributions  are not disjoint: 1.0\% of field pairs at solar metallicity have $\chi^2$ differences as small as the median $\chi^2$ among intra-cluster pairs; these stars are {\it doppelganger}.}
\label{fig:within}
\vspace{10pt}
\end{figure*}

To explore the similarity (over all or many elements) in abundance space for stars within clusters and among field stars, one may be tempted to devise a simple 
pseudo-cartesian measure of distance in abundance space, such as $D_{nn'}^2 = \sum_{i=1}^{I} \left ( x_{ni} - x_{n'i}\right )^2$,
and then compare the distributions $p(D^{cluster}_{nn'})$ and $p(D^{field}_{nn'})$. 
However, with imperfect measurements $x_{ni}$, the use and interpretation of scalar distance measurements in a high-dimensional space is complex and depends strongly on the choices of prior assumptions,
whenever the distances become comparable to the measurement uncertainties, as is the case here;
we sketch this issue out in the Appendix. But it is straightforward to assess, via $\chi^2$, how likely the abundance measurements of a pair of stars would be, if the true abundances were identical: 

\begin{equation}
\chi_{nn'}^2 = \sum_{i=1}^{I} \frac{[x_{ni} - x_{n'i}]^2}{\sigma_{ni}^2 + \sigma_{n'i}^2}.
\end{equation}

The two star indices are $n$ and $n'$, and each star $n$ has a measurement $x_{ni}$  and each star $n'$ has a measurement $x_{n'i}$, for elements $i$=1 to I, with uncertainty $\sigma_{ni}$ and $\sigma_{n'i}$, respectively. We adopted the median-rescaled uncertainties from
Section \ref{uncertainty_rescale} here. For each star we determine the neighbor most likely to have identical abundances by minimizing ${\chi_{nn'}^2}$ 

This $\chi_{nn'}^2$ metric is then calculated for star pairs within the same cluster as well as for the stars pairs drawn from a random sample of 2000 field stars, which have analogously determined abundances (and uncertainties).

\begin{figure}
\centering
\includegraphics[scale=0.28]{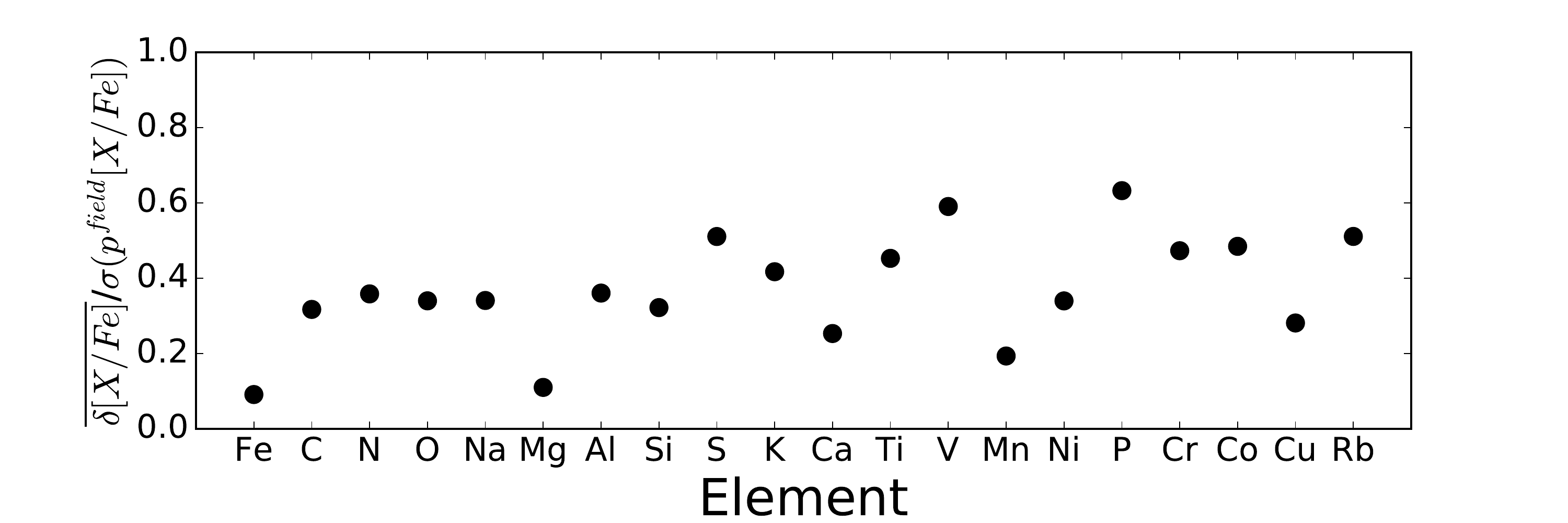} 
  \caption{The ratio of the mean measurement error for each element and the 1-$\sigma$ dispersion of that element from the field sample. This ratio is a measure of the discriminating power of each of the elements in determining chemical dissimilarity. A ratio of 1 indicates that the typical error bar on the measurement is the same as the spread in that abundance in the field: elements with a high ratio have the least discriminating power, elements with a low ratio, the most. All elements are [X/Fe] except Fe, which is [Fe/H]. }
\label{fig:power}
\vspace{10pt}
\end{figure}

The field stars were chosen to have a logg $<$ 3.9 (giants) and a SNR $>$ 200 to be comparably high in SNR to the cluster sample. The results for all pairs, i.e. the distributions $p^{cluster}_{pair}(\chi^2)$ and $p^{field}_{pair}(\chi^2)$, are shown in the top panel of Figure \ref{fig:within}. For these pairs we restricted comparisons to within $\Delta$\teff\ $<$ 315K and $\Delta$\logg\ $<$ 0.7, to guard against differences in these ``nuisance labels" leading to systematic abundance differences; we found that beyond these limits the $\chi^2$-distance in abundance space is correlated with these parameter differences. The $p_{pair}(\chi^2)$ for pairs from the same cluster are shown in black and pairs of 2000 random field stars are shown in the red (dashed line). 

Figure \ref{fig:within} demonstrates that in 20-element abundance space stars within clusters are, unsurprisingly, chemically far more similar than star pairs among the field sample. For pairs within a cluster, $p^{cluster}_{pair}(\chi^2)$ peaks at $\sim 20$, as expected if {\it all} stars in a cluster had identical 
abundances in {\it all} elements; however, there is a significant tail of $p^{intra}_{pair}(\chi^2)$ towards
substantively higher values of $\chi^2$, implying that {\sl some} stars differ in {\sl some} elements even within
a cluster. Note that we have included the elements $C$ and $N$ in this comparison, even though their photospheric abundances get altered by dredge-up, to a degree that depends in giants on the mass or age. For those elements, similarity implies a combination of near-identical birth-material and age, as expected for open clusters.

For field star pairs $p^{field}_{pair}(\chi^2)$ is far broader, which is unsurprising, as members of a random field star pair will usually differ even in their most elementary abundance, [Fe/H]. The vast 
majority of the $p^{field}_{pair}(\chi^2)$ values lie at $\chi^2$ values far in excess of  $\chi^2 \sim 20$. However, there is a small fraction of field pairs ($\sim 1-2\%$) whose 20-element abundances are indistinguishable $\chi^2 \le 25$ despite the 0.02-0.03 dex precision that \apogee\ affords for individual elements.

An obvious next question to ask is whether the differences between $p^{cluster}_{pair}(\chi^2)$ 
and $p^{field}_{pair}(\chi^2)$, shown in Figure \ref{fig:within} are primarily driven by 
differences in the basic [Fe/H], rather than the high-dimensional [X/Fe]. To do this, we 
drew up the distributions $p^{cluster}_{pair}(\chi^2)$ and $p^{field}_{pair}(\chi^2)$, but restricted to solar [Fe/H]. For the intra-cluster pairs we consider only two clusters, M67
([Fe/H]=0.0) and NGC6819 ([Fe/H]=0.03); for this second selection of field pairs, this meant we restricted the sample to red giant stars with [Fe/H]$=0.00\pm0.02$. A total of 1054 solar metallicity field giants (again selected with SNR $>$ 200) were used for this comparison. The resulting distributions are shown in the bottom panel of Figure \ref{fig:within}.
This Figure shows that the two distributions (intra-cluster and field pairs) remain distinctly different: there is valuable discriminating information in the [X/Fe]. However, considering
{\it a priori} only pairs of the same metallicity naturally increases the overlap between the distributions substantially.

In Figure \ref{fig:within} there are many more pairs (especially intra-cluster pairs) at very small values of $\chi^2$, far more than expected from the chi-squared distribution with 19 or 18 degrees of freedom.
This could have multiple origins, related to our chemical-abundance uncertainty model.
Our chemical-space uncertainty analysis is fundamentally empirical; it presumes that all stars are comparable in their noise properties.
If instead there are variations in the uncertainties, or if the chemical-abundance estimate noise is very non-Gaussian, the distribution of $\chi^2$ values can depart strongly from a chi-squared distribution.
Presumably we are seeing this effect.
In addition, the intra-cluster pair distribution contains a far larger fraction of very small $\chi^2$ values compared to  the field pair distribution. This suggests that there are at least some stars for which the chemical abundances are measured with true uncertainties that are much smaller than our baseline estimates.
That is, we may be measuring things more precisely than what is implied by our current (relatively conservative) uncertainty analysis.

\section{The Fraction of Galactic Doppelgangers}

Figure \ref{fig:within} characterizes the distribution of the abundance similarity of stars that are known to be born together $p^{cluster}_{pair}(\chi^2)$, compared to that of random field stars, $p^{field}_{pair}(\chi^2)$. From this, we can determine the relative fraction of stars that appear as chemically similar as cluster stars do to one another, yet are not born in the same environment, so called \textit{doppelgangers}. 

For the field, 0.3\% of random pairs have a $\chi^2$ difference that is as small as the median $\chi^2$ among intra-cluster pairs. An additional constraint for stars that are born together is that these stars must have the same metallicity.  Therefore, to assess the implications for the viability of strict chemical tagging, we should consider the abundance similarity of stars at a single metallicity, as shown in the bottom panel of Figure \ref{fig:within}.  For the stars at a single (solar) metallicity, \textbf{1.0\%} of field pairs have $\chi^2$ differences as small as the median $\chi^2$ among intra-cluster pairs; these stars are doppelganger.

In examining the similarity of the field pairs at the same metallicity of [Fe/H] = 0, probable new members of two clusters were discovered, as detailed in Section 6. Although negligible in relative number of all the pair combinations, these were removed from the determination of the doppelganger rate, calculated above (to be 1.0\% at [Fe/H] = 0). The new probable members identified  demonstrates that, with the $\chi^2$ pair similarity analysis, we have developed an effective technique to identify new cluster members, when combined with radial velocity measurements. 

\section{Identification of new cluster members}

We examined all `sibling' field pairs that are as chemically similar as cluster pairs, to ensure they are doppelgangers and not true cluster members that are simply not included in our list of analyzed clusters.  From this, we found four new potential cluster members (all with $\chi^2$ $<$ 7); one pair associated with M67 and one pair associated with NGC7789. 

We found these pairs only by using their chemical similarity to each other (as similar as that of the intra-cluster pairs) and then by observing that they had a common velocity and were in proximity to open clusters of the same [Fe/H] and mean radial velocity as the cluster. 

\begin{figure*}
\centering
\includegraphics[scale=0.4]{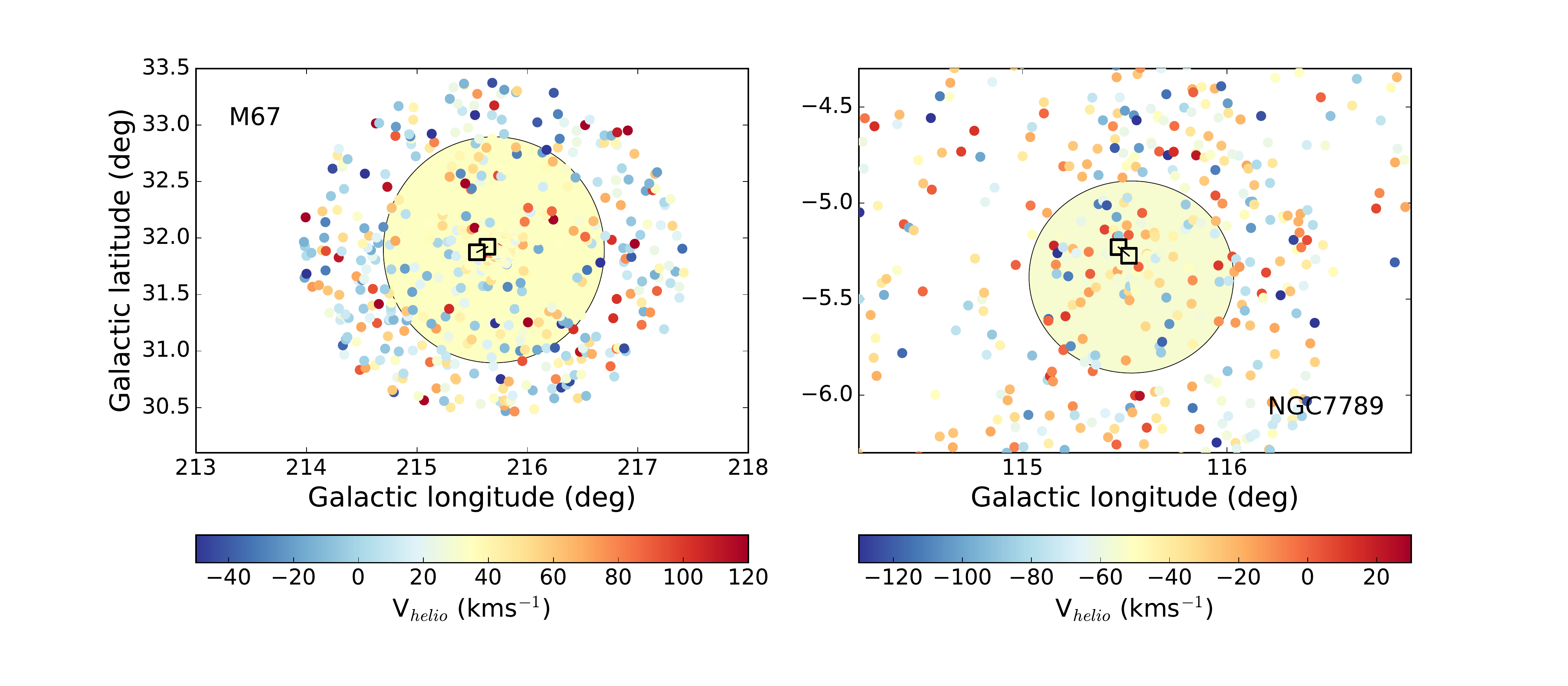} 
  \caption{The clusters M67 and NGC7789, represented by circles of the size of their tidal radius, the new potential member pairs identified (colored squares joined with lines) and surrounding field stars (small colored circles), all colored by heliocentric velocity (showing the mean heliocentric velocity for the clusters). The new stars identified have the same radial velocities as well as chemical abundances, for most elements, as the respective clusters (see Figure \ref{fig:C2a}).}
\label{fig:C}
\end{figure*}

The properties of the two newly identified member pairs from the field are summarised in Table 3 (Appendix).  The positions of these stars and their radial velocities along with the cluster centre and surrounding field stars are shown coloured by radial velocity in Figure \ref{fig:C}. The stars associated with NGC7789 have velocities of -53.6 and -54.9 km$s^{-1}$ and the stars we associate with M67 have velocities of 34.2 and 34.1 km$s^{-1}$. These velocity measurements are all within the respective cluster velocity dispersions: the mean velocity and velocity dispersion of M67 is 33.6 $\pm$ 0.8 kms$^{-1}$ \citep{Geller2015}. The mean velocity and velocity dispersion of NGC7789 is  -54.9 $\pm$ 0.9 kms$^{-1}$ \citep{Gim1998}.

These pairs were excluded from our doppelganger rate determination in Section 5, as they are potentially actual cluster members. We emphasize that these were \textit{not} identified as members of the clusters via a test of the similarity of their 19 abundances to the clusters themselves.  

\begin{figure*}
\centering
\includegraphics[scale=0.4]{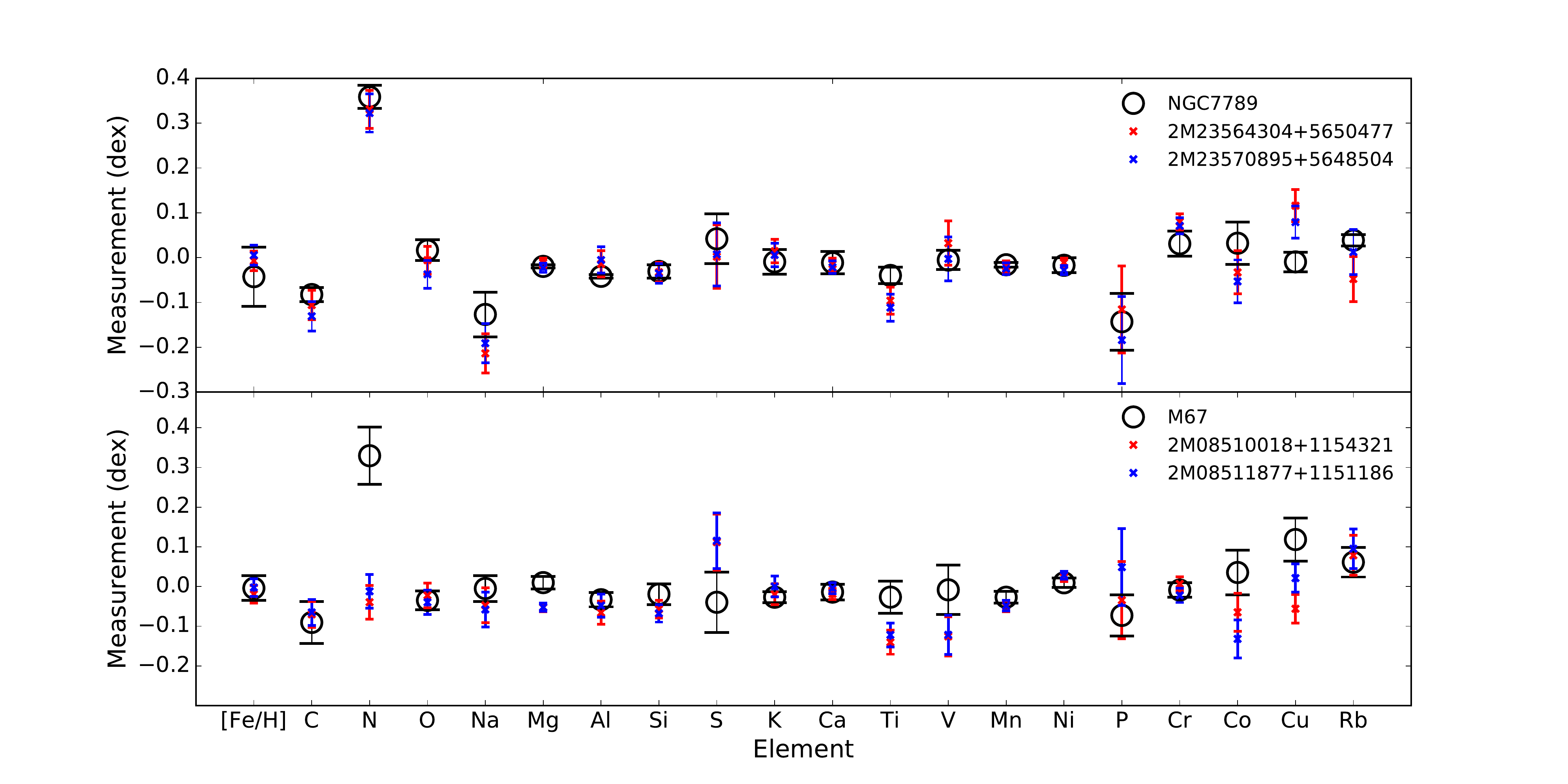} 
  \caption{The mean abundance measurements and corresponding error bars for the clusters NGC7789 (top) and M67 (bottom) shown in the open black circles with the potential new members identified from the field using the pair abundance similarity metric, shown in the colored crosses. The 2MASS IDs of the new member stars are listed. The [N/Fe] measurements for the pair of the stars identified as M67 members are much lower than the mean abundance of the cluster. This difference is likely a consequence of CNO material mixing as stars move up the giant branch \citep[e.g.][]{Holtzman2015} and that these stars, at the base of the giant branch (see Figure \ref{fig:C_tg}), have not yet experienced the first dredge up \citep{Martig2016}. These two stars have hotter temperatures than all of the other observed cluster stars. The stars are outside of the 1-$\sigma$ mean error bars for the cluster in some cases (e.g. [Mg/Fe], [S/Fe], [Cu/Fe]); cases which do not have an astrophysical explanation. This may be a consequence of small residual systematic temperature trends with these measured elements. Note that the pair abundance similarity determinations are restricted to stars within a narrow range of temperature and gravity, but the pairs themselves in comparison to the cluster stars can have dissimilar temperatures. }
\label{fig:C2a}
\end{figure*}

These pairs have abundance measurements that are generally consistent with those of the clusters with some exceptions. These exceptions in particular, in the case of M67, lead to a high $\chi^2$ ($>$ 20) between the potential new members of M67 and the known M67 member stars. \textit{Based on a comparison of the abundances of these two stars with the known M67 stars alone, these would not be considered members, under the assumption of near zero intrinsic abundance spread}. These differences however, may be associated with temperature and \logg\ systematics -- including those of an astrophysical origin (see the Appendix). The abundances in all 20 elements of the potential new member stars are shown with the mean values for the M67 and NGC7789 clusters in Figure \ref{fig:C2a}. The mean \teff\ and \logg\ of the M67 cluster stars is 4770K and 2.8 dex, respectively. The pair of stars we associate with M67 are hotter and have higher gravities, at $\sim$ 5200K and 3.7 dex (see Table 3 and Figure C1 in the Appendix). Although the high $\chi^2$ is particularly driven by a few outlying elements in particular (e.g. [N/Fe] and [Mg/Fe]), even with the largest outliers removed from the $\chi^2$ comparison, the new potential members remain at a larger $\chi^2$ distance from the M67 members compared to the intra-cluster pair comparison. For the pair associated with NGC7789, with temperature and \logg\ values around 4900K and 2.7 dex, (compared to the cluster mean of 4600 K and 2.3 dex), three of the NGC7789 cluster stars are within $\chi^2$ $<$ 19 of these potential new members.  The proper motions, included in Table 3, are consistent with these four stars being open cluster members.

\section{Discussion and Conclusion}

In this paper we set out to quantify the similarity among Galactic disk stars with respect to their 
detailed photospheric element abundances, to clarify the prospects of chemical tagging with data of \apogee 's quality.

For such an analysis we could draw on a unique abundance data set, derived by applying a modified
version of \tc\ to \apogee\ spectra to assure maximal abundance precision and well-understood uncertainties.
This re-analysis allowed for simultaneous fitting the spectral line-spread-function. Thus, eliminating a likely a source of 
subtle but correlated systematic errors in the abundance estimates. The quantification of the abundance uncertainties
included SNR-dependent cross-validation and an error-rescaling for some elements, exploiting the fact that
abundance estimates for member-stars of open clusters should on average not differ by less than their uncertainties.

This left us with a set of \apogee -derived abundances that is unprecedented 
in its (pertinent) combination of quality, quantity and homogeneity. We only considered stars with very similar 
stellar parameters, giants in a restricted $T_{eff}$ and log~g range; 
we have 90 spectra across seven open clusters and thousands of ``field stars'' with the same data, 
SNR and data analysis. For all these stars we have individual abundances for 20 elements 
(Fe, C, N, O, Na, Mg, Al, Si, S, K, Ca, Ti, V, Mn, Ni, P, Cr, Co, Cu, Rb), with a median precision of 0.03 dex. We characterize these abundances
by their values [X/Fe], except for iron where we use [Fe/H].

On this basis, we undertook an extensive analysis of the chemical (in-)homogeneity of stars within an open cluster, 
testing the standard assumption that such stellar ``birth siblings''
were formed at the same time from chemically homogeneous material. This should result in identical element abundances,
including those elements (e.g. $C$ and $N$), whose photospheric abundance has been altered in giants by mass/age-dependent
dredge-up. We confirm that the abundances of cluster member stars typically agree with each other within 
their small (0.03 dex) uncertainties, as had been found before (see Section 1). Moving beyond, we explicitly determined the
{\it intrinsic} dispersion in these abundances, or derived upper limits for them. We found that 
(at least in the best-sampled clusters) the abundance dispersions are essentially zero, 
with typical upper limits of $\sim 0.03$~dex. However, there are exceptions: some elements in some clusters show
small but significant dispersion, attributable to a subset of the member stars. 

To compare the abundance similarity of stars with a known common birth origin (cluster members) to 
the mutual similarity of field stars, we compared the abundances among pairs of stars. It was tempting to 
define a distance measure in the high-dimensional space of element abundances; but the ``curse of dimensionality''
makes distance in a 20-dimensional space very dependent on prior assumptions about distance distributions, once the distances
become comparable to the measurement uncertainties. Therefore, we resorted to a simple $\chi^2$ statistic, 
$p_{pair}(\chi^2)$, quantifying the likelihood of the 20 observed abundance (differences) 
for any pair of stars, assuming their intrinsic abundances were identical.

When we construct $p^{cluster}_{pair}(\chi^2)$ for intra-cluster pairs of stars, we find that its median is similar to 
$N_{elements}\approx 20$, again illustrating the level of chemical homogeneity within clusters. However, about 20\% of intra-cluster star pairs have $\chi^2\ge 40$ indicating significant abundance differences in one or more elements.
We find open clusters to be mostly, but not exclusively homogeneous.

When we construct $p^{field}_{pair}(\chi^2)$ for pairs of field stars, it looks overall very different. 
The vast majority of the support of this distribution lies at $\chi^2\ge 40$: most pairs of field stars can be clearly recognized as having differing abundances. In particular, 99.7\% of the field pairs have $\chi^2$ in excess of the median $\chi^2$ for intra-cluster pairs. Part of why $p^{field}_{pair}(\chi^2)$ looks so different from
$p^{cluster}_{pair}(\chi^2)$ is of course that the Galactic (thin) disk has a metallicity spread of about 1~dex;
but considering the analogous distributions restricted to solar metallicity, shows that the remaining abundances
still have great discriminating power. Most commonly the abundances of field pairs -- at the same [Fe/H] (e.g. $|$[Fe/H]$|$ $<$ 0.02) -- are inconsistent with being identical; 99\% of field star pairs of the same metallicity appear chemically non-identical on the basis of their other element abundances with \apogee\ data. 

Of course, the above result implies in turn that 0.30\% of random pairs among Galactic disk have indistinguishable abundances,
even with typical 0.03~dex measurement precision for $\sim 20$ elements that \apogee\ data afford; this has remarkable consequences. 
Such pairs are much more common than any plausible incidence of common-birth-site (former cluster), $10^{-4~...~-6}$.
Therefore most of them cannot be birth siblings, but mere {\it doppelganger}: stars not immediately related by birth, yet looking near-perfectly alike in their abundance patterns. This rate of field star {\it doppelganger} has not been
quantified before, and has important implications for chemical tagging. Note that this rate estimate applies to stars with typical Galactic disk abundances, {$-0.7\le$[Fe/H]$\le 0.3$} and $[\alpha$/Fe] $\sim$ solar, where abundance space is most densely populated with stars. In the metal poor old disk or the halo, {\it doppelganger} may be much more lower \citep{Ting2015}, as abundances show greater diversity, 
and therefore abundance space is more sparsely populated.
It will be interesting to explore whether the inclusion of extensive sets of $s$- and $r$-process elements (e.g. from Galah) will change this picture qualitatively.

It is also worth noting that the $\chi^2$ statistic that we use here -- while extremely straightforward -- is unlikely the optimal way to discriminate cluster pairs from field pairs. In particular, there are elements (foremost [Fe/H] and $\alpha$-elements) in which the abundance variation among field pairs is large, compared to our measurement precision. Conversely, there may be elements (when viewed as [X/Fe]), where even random field pairs are unlikely to differ by more than the measurement uncertainties (see Figure \ref{fig:power}). Those elements are then rather uninformative when it comes to discriminating the two hypothesis, and will mostly add variance in the $\chi^2$ sum. This suggests that a sum of the individual element differences, without weighting and covariance terms, is likely sub-optimal. However, tests on our current set of elements, taking only a subset of the most informative elements for our $\chi^2$ calculation indicates there is no gain expected from removing the least informative elements. A rigorous development of a near-optimal statistic to discriminate birth-siblings from field pairs (i.e. minimizing the doppelganger rate) requires a full characterization of the (error-deconvolved) 20-dimensional abundance distribution of the field \citep[see also][]{Ting2015}. This is beyond the scope of this paper. However, simple experiments to restrict the $\chi^2$ sum of the abundance differences to subsets of the most discriminating elements, showed that the basic picture drawn up here is unlikely to change; the existence of an important doppelganger population in \apogee -like data  is not a consequence of the sub-optimal $\chi^2$ statistic.

To summarize the assessment of abundance ($\chi^2$) similarities or differences, 
with 20 abundances at a typical 0.03~dex precision level at hand:
most - but not all -- pairs of stars in clusters are chemically alike; most -- but not all -- pairs of field stars  are chemically different.

In conclusion, we can now discuss what these findings mean for ``chemical tagging''. Such studies of chemical similarity
among stars in the context of Milky Way evolution have claimed two main goals. One is an exploration of the 
Galactic history of chemically homogeneous birth sites of stars, i.e. the history of the (now disrupted) cluster mass function \citep{Ting2016}. This ``strict chemical tagging" would require attributing a common birth origin to, say, pairs of stars with considerable confidence, on the basis of abundances alone. Even in optimistic scenarios for the main stellar disk of the Milky Way, only $10^{-4.5}$ random star pairs would be such birth siblings (our doppelganger measurement is equivalent to $10^{-2}$).
The other goal is the empirical derivation of a successively more detailed chemo-orbital distribution of stars in the Galaxy: stars of which abundances are on which orbits. Even considering only [Fe/H] and [$\alpha$/Fe], it is well known that 
the Galactic disk structure varies distinctly \citep{Bovy2012, Hayden2014}. Therefore, more intricate patterns, involving more abundances and ages could and should provide better constraints on radial migration or heating.
Such ``broad chemical tagging" (or chemical labeling) studies take detailed abundances foremost as a lifelong label of stars, 
independent of orbital phase, without making explicit reference to a common birth site among pairs or groups of stars.

Our results presented here, in particular the significant incidence of {\it doppelganger} stars in the field imply that
strict chemical tagging is presumably not possible, certainly not for the Milky Way's main stellar disk components, even with the data quality level presented here. When considering
whether a pair of stars was likely born in the same site at the same time, one must consider not only the
data likelihoods presented above, but also the prior expectation for ``sibling'' or {\it doppelganger}, $p_{sibl}$
and $p_{dg}$, respectively. Then we have
\begin{equation}
\frac{p(sibl~|~data)}{p(dg~|~data)} = \frac{p(data~|~sibl)}{p(data~|~dg)}\times\frac{p_{prior}(sibl)}{p_{prior}(dg)},
\end{equation}
where $$p(data | sibl)/p(data | dg)\approx p^{cluster}_{pair}(data)/p^{field}_{pair}(data)$$ from Figure \ref{fig:within},
and the ratio of priors is $p_{prior}(sibl)/p_{prior}(dg)=10^{-4~...~-6}$. As Figure \ref{fig:within} shows, 
$p^{cluster}_{pair}(data)/p^{field}_{pair}(data)$ have basically the same support and that the ratio of priors is so small,
it seems hard to imagine to get $\frac{p(sibl~|~data)}{p(dg~|~data)} \ge 1$.

While this analysis may dampen the prospects for strict chemical tagging, it is good news for detailed chemical labeling;
the analysis shows that the Galactic stellar disk population can be ``sliced'' into much more detailed abundance-based ways,
which will provide new, and probably powerful, ways to constrain the formation of the Milky Way. The $C$ and $N$-based
age estimates are just one recent example of this approach. Furthermore, combining velocity information (and eventually full dynamics) with the abundance analysis demonstrates promising prospects for identification of cluster members. 

\section{Acknowledgements}
The authors thank Andy Gould for useful discussions. The authors thank Baitan Tang for a review of the manuscript. 

M. Ness and H.-W. Rix acknowledge funding from the European Research Council under the
European Union's Seventh Framework Programme (FP 7) ERC Advanced Grant Agreement n. [321035]. H.-W. Rix acknowledges support of the Miller Institute at UC Berkeley through a visiting professorship  during the completion of this work.

DWH was partially supported by the NSF (grants IIS-1124794 and AST-1517237), NASA (grant NNX12AI50G), and the Moore-Sloan Data Science Environment at NYU.

D.G. and D.M. gratefully acknowledge support from the BASAL Centro de Astrofsica y Tecnologias Afines (CATA) grant PFB-06/2007. D.M. is also supported by  the Ministry for the Economy, Development and Tourism, Programa Iniciativa Cientifica Milenio grant IC120009, awarded to the Millennium Institute of Astrophysics (MAS), and by FONDECYT No. 1130196

PMF gratefully acknowledges support from the National Science Foundation through award AST–1311835

Funding for the Sloan Digital Sky Survey IV has been provided by the
Alfred P. Sloan Foundation, the U.S. Department of Energy Office of
Science, and the Participating Institutions. SDSS acknowledges
support and resources from the Center for High-Performance Computing at
the University of Utah. The SDSS web site is www.sdss.org.

SDSS is managed by the Astrophysical Research Consortium for the Participating Institutions of the SDSS Collaboration including the Brazilian Participation Group, the Carnegie Institution for Science, Carnegie Mellon University, the Chilean Participation Group, the French Participation Group, Harvard-Smithsonian Center for Astrophysics, Instituto de Astrofisica de Canarias, The Johns Hopkins University, Kavli Institute for the Physics and Mathematics of the Universe (IPMU) / University of Tokyo, Lawrence Berkeley National Laboratory, Leibniz Institut für Astrophysik Potsdam (AIP), Max-Planck-Institut fur Astronomie (MPIA Heidelberg), Max-Planck-Institut fur Astrophysik (MPA Garching), Max-Planck-Institut fur Extraterrestrische Physik (MPE), National Astronomical Observatories of China, New Mexico State University, New York University, University of Notre Dame, Observatório Nacional / MCTI, The Ohio State University, Pennsylvania State University, Shanghai Astronomical Observatory, United Kingdom Participation Group, Universidad Nacional Autonoma de México, University of Arizona, University of Colorado Boulder, University of Oxford, University of Portsmouth, University of Utah, University of Virginia, University of Washington, University of Wisconsin, Vanderbilt University, and Yale University.

\appendix

\section{Distances in High-Dimensional Spaces}\label{sec:appendix}

In the context of chemical tagging it is tempting to assign a scalar or vector measure of abundance-space distance, $D$ (or $\vec{D}$) to pairs of stars.
Such a measure of distance would have to be derived  from  abundance measurements with uncertainties. In this  Appendix we show why the interpretation of such distance measures becomes problematic in high-dimensional  spaces, as soon as the true or presumed distances become comparable to the measurement uncertainties. This is why in the analysis presented in this paper, we restrict ourselves  to asking how likely the abundance data on a pair of stars are (in the $\chi^2$-sense) if their abundances were identical.

Let us first consider how high-dimensional the space of abundances is; already here is no simple unique answer. Several measures of dimensionality may play a role here: in principle, the dimensionality of abundance space is the length of the periodic system (and then, there are isotopes); in practice, measurements only provide constraints on a subset of elements, between 1 and 35 elements;  finally, there is the question of how many astrophysically non-degenerate dimensions abundance space has.

In principle, it is straightforward to assign a metric to abundance space,
e.g. some single Cartesian measure of distance 
between two points in $L$-dimensional abundance space: $D_{nn'}^2\equiv\sum_{i=1}^I D_i^2$, with $D_i\equiv [X_{n,i}/H]-[X_{n',i}/H]$.  
However, in practice we would need to make a probabilistic statement  about the abundance-space distance between two stars, given a set of abundance measurements for those two stars. To get a {\it pdf} for any measure of distance, one would have to specify  a prior expectation for that distance 
measure, which  then gets modified by the data likelihood.
But if one wants to have a {\it pdf} for a simple scalar/1D distance measure in light
of independent measurements in a high-dimensional abundance space, it appears difficult to avoid that the resulting  $p_{pdf}(D_{scalar})$ depends very strongly on the priors, if the  intrinsic distance becomes comparable to the measurement uncertainties; this is the pertinent regime for chemical tagging.

To illustrate this, we will spell out the case for a one-dimensional abundance measurement  specifically, and then discuss how to generalize them. 
For a pair of stars we have measurements of their abundances  \xone and \xtwo , with their measurement uncertainties  \done and \dtwo .  We  presume that their true abundances can 
be characterized by \xbar\ and $D$, as $\xbar - D/2$ and $\xbar + D/2$, with $D=2d$.
We then get 
\begin{equation*}
p_{pdf}(D,\xbar \given \xone , \xtwo ,\done , \dtwo ) 
\propto p_L(\xone , \xtwo\given D,\xbar,\done , \dtwo)\cdot p_{prior}(D,\xbar).
\end{equation*}
We can now spell out the data likelihood
\begin{multline*}
 p_L(\xone , \xtwo\given D,\xbar,\done , \dtwo)=\\
 \frac{1/2}{2\pi \done \dtwo}
 \left ( \exp -\left [ \frac{(\xone-\xbar+D/2)^2}{2\done^2}+
 \frac{(\xtwo-\xbar-D/2)^2}{2\dtwo^2}\right ]
 +\exp -\left [ \frac{(\xone-\xbar-D/2)^2}{2\done^2}+\frac{(\xtwo-\xbar+D/2)^2}{2\dtwo^2} \right ]\right ) ,
\end{multline*}
when one integrates over \xbar\ (with a flat prior in \xbar ) and simplifies this becomes
\begin{equation}\label{eq:distance}
 p_L(\xone - \xtwo\given D, \done , \dtwo)=
 \frac{1}{\sqrt{2\pi (\done^2 + \dtwo^2)}}
 \exp{\left (- \frac{D^2+(\xone-\xtwo)^2}{2(\done^2+\dtwo^2)}\right )} \times
\cosh{\left ( \frac{D(\xone-\xtwo)}{\done^2+\dtwo^2}\right )}.
\end{equation}
For the hypothesis that the abundances are identical, i.e. $D\equiv0$, this becomes:
\begin{equation*}
 p_{L,D\equiv 0}(\xone - \xtwo\given\done , \dtwo)\propto 
 \frac{1}{\sqrt{2\pi (\done^2 + \dtwo^2)}}
 \exp{\left (- \frac{(\xone-\xtwo)^2}{2(\done^2+\dtwo^2)}\right )} ,
\end{equation*}
corresponding to the simple 
$\chi^2$-expression from Eq. 2, describing the likelihood of the data under the hypothesis that  $D=d\equiv0$. 

An extension of Eq.\ref{eq:distance} to the $L$-dimensional case, say $L=10$, brings us to a well-established conundrum. If we stick to the seemingly natural prior
$p_{prior}(D_l)=\mathcal{H}(D_l)$ for all $l$, then this implies as a prior 
for the $L$-distance $D_L$: $p_{prior}(D_L)\propto (D_L)^L$.
The same problem arises for any prior  $p_{prior}(D_l)$ that is flat for small 
$|D_l|$.  Like a $\chi^2$ distribution,  the {\it inferred} most likely (scalar) distance $D^2$ would always tend to the squared sum of the measurement uncertainties.  

\section{New Cluster Members}
%
\begin{deluxetable}{lccccccc}
\tablecaption{ Newly identified NGC7789 and M67 Cluster Members from the Field \label{t:tab3}}
\tablecolumns{8}
\tablenum{3}
\tablewidth{0pt}
\tablehead{
\colhead{No.} & \colhead{2MASS ID} & \colhead{V$_{helio}$} & \colhead{Distance from cluster centre}  & \colhead{Teff} & \colhead{log g} & \colhead{PMRA} & \colhead{PMDEC} \\
\colhead{} & \colhead{} &   \colhead{(km$s^{-1}$)} & \colhead{(deg)} & \colhead{(K)} & \colhead{(dex)} & \colhead{masyr$^{-1}$} & \colhead{masyr$^{-1}$} \\
}
\startdata
1 & 2M23564304+5650477 &   -53.63 & 0.1 & 4890 & 2.7 & -4.7 $\pm$ 5.3 & -1.6 $\pm$ 5.3 \\
 & 2M23570895+5648504 &    -54.90 & 0.17 & 4927 & 2.7 & 0.1 $\pm$ 3.9 & 2.1 $\pm$ 3.9 \\
\hline
2 & 2M08510018+1154321 &     34.2 & 0.15 & 5202 & 3.7 & -7.4 $\pm$ 1.0 & -5.4 $\pm$ 1.0 \\
 & 2M08511877+1151186 &     34.1 & 0.06 & 5161 & 3.7 & -6.9 $\pm$ 1.1 & -6.2 $\pm$ 1.1 \\
\hline
\enddata
\tablecomments{Identified additional (pair) members of the clusters NGC7789 (pair 1) and M67 (pair 2). The proper motions are sourced from the PPMXL catalogue \citep{Roeser2010}. The Proper motion of M67 is (PMRA, PMDEC) = (-7.64 $\pm$ 0.07, -5.98 $\pm$ 0.07) masyr$^{-1}$ \citep{Gao2016} and of NGC7789 is  (PMRA, PMDEC) = (-2.2 $\pm$ 0.22, -1.1 $\pm$ 0.22) masyr$^{-1}$. }
\end{deluxetable}

\section{Mixing of Carbon and Nitrogen along the giant branch}

\begin{figure*}[h]
\centering
\includegraphics[scale=0.25]{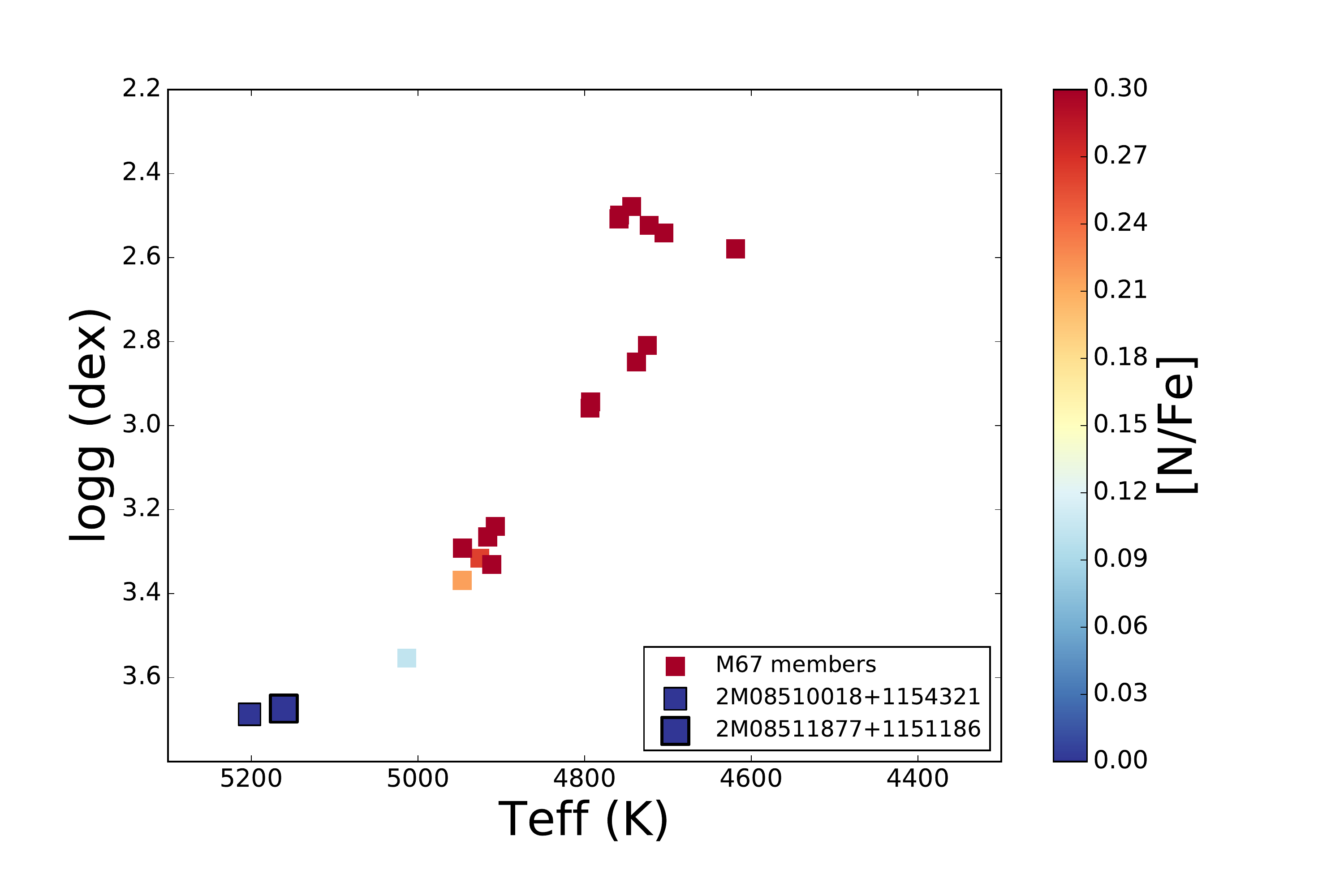} 
\includegraphics[scale=0.25]{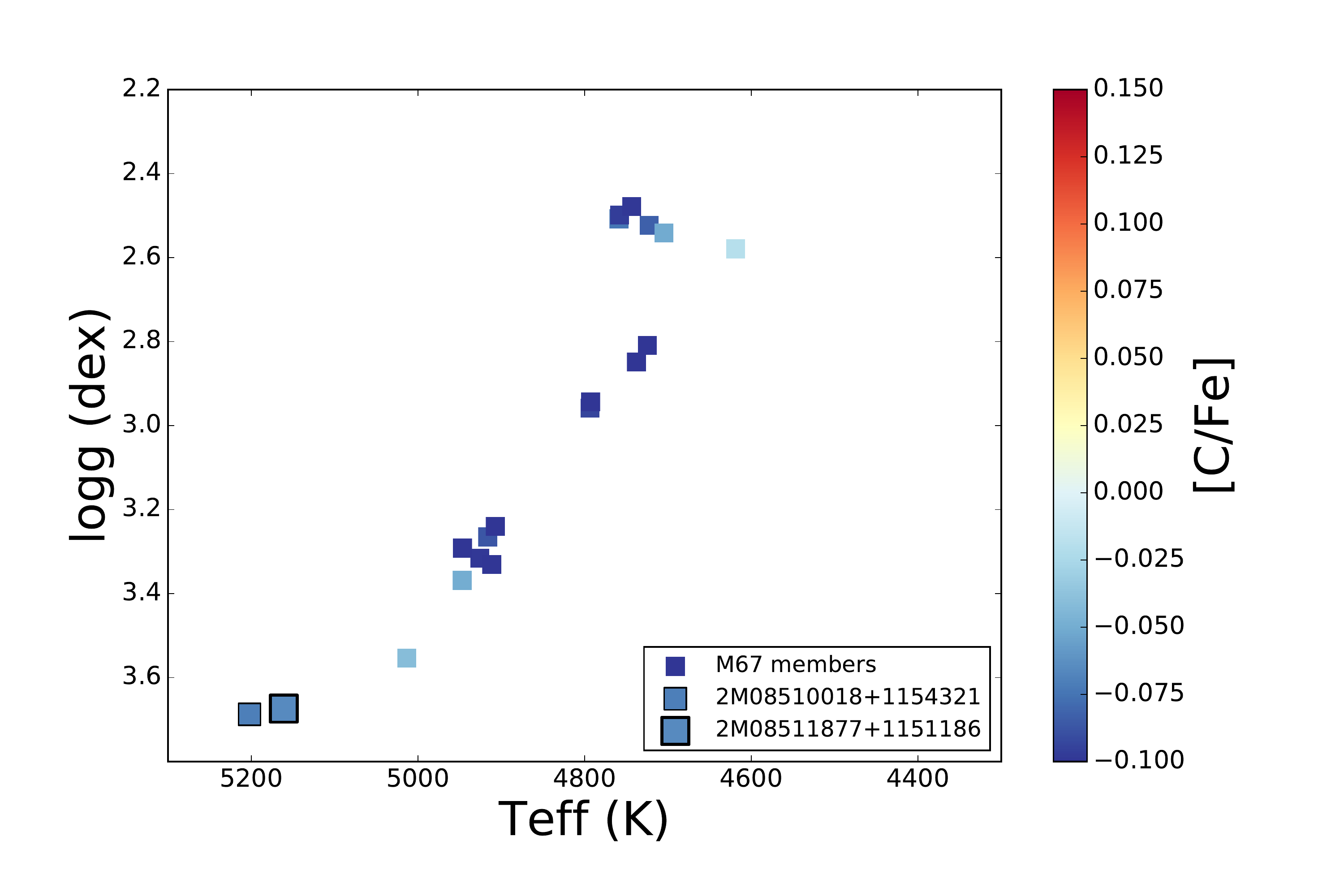} 
  \caption{The temperature and \logg\ values of the M67, including both already determined APOGEE members used for our analysis and the new members identified by measuring abundance similarities between field pairs, colored by their [N/Fe] (left)  and [C/Fe] values (right). The [C/Fe] and [N/Fe] ratios in the stellar atmosphere change up the giant branch as there is mixing from the interior of the star. The two newly identified M67 stars, which are outliers in a few of the abundances compared to the mean M67 cluster measurements, most notably for [N/Fe], are at the base of the giant branch, at hotter temperatures and higher gravities than the stars in the cluster from which mean results have been determined. These two stars would not have yet undergone the first dredge up. }
\label{fig:C_tg}
\end{figure*}

Figure \ref{fig:C_tg} shows both the known cluster stars and our new members in the \teff\ and \logg\ plane for M67. For  the pair comparisons, the stars are restricted to being near in \teff\ and \logg. However for the comparison shown in Figure \ref{fig:C2a}, while the pairs are themselves similar, these are not required to be similar to the cluster stars from which the mean abundance values are determined.  There are stars that are newly identified members at the hot end of the giant branch and these stars have a [N/Fe] measurement that falls outside the 1-$\sigma$ mean cluster measurements (calculated from stars at intermediate \teff\ and \logg\ values along the giant branch). There are also small discrepancies in a few of the elements in M67 for the hot stars at the base of the giant branch compared to the mean abundance of M67 stars and this is also likely due to small systematic dependencies in temperature from the stellar model abundance determinations propagated by The Cannon (see the caption of Figure \ref{fig:C2a}).

\end{document}